\documentclass[twocolumn]{aastex62}

\newcommand{\htwo}{\mbox{H$_2$}}

\newcommand{\hi}{\mbox{H{\small I}}}
\newcommand{\cii}{\mbox{[C{\small II}]}}
\newcommand{\nii}{\mbox{[N{\small II}]}}

\newcommand{\hii}{\mbox{H{\small II}}}

\newcommand{\msun}{\mbox{\,M$_{\odot}$}}

\newcommand{\kms}{km\,s$^{-1}$}

\newcommand{\density}{$\mathrm{cm^{-3}}$}
\newcommand{\fneut}{\mbox{$\mathrm{f_{\rm neutral}}$}}
\newcommand{\sSFR}{\mbox{$\mathrm{\Sigma_{SFR}}$}}
\newcommand{\fmol}{\mbox{$\mathrm{f_{\rm mol}}$}}
\newcommand{\fatomic}{\mbox{$\mathrm{f_{\rm atomic}}$}}
\newcommand{\gc}{\mbox{$\mathrm{R/R_{25}}$}}
\newcommand{\SFRcut}{\mbox{$\sSFR = 6 \times 10^{-2} \rm \ \msun\ \ yr^{-1} \ kpc^{-2}$}}
\newcommand{\changes}{}

\usepackage{amsmath}
\usepackage{hyperref}
\usepackage[inline]{enumitem}

\graphicspath{{./}{figures/}}

\accepted{April 26, 2021}
\submitjournal{ApJ}

\shortauthors{Tarantino et al.}

\begin{document}

\title{Characterizing the Multi-Phase Origin of \cii \ Emission in M101 and NGC 6946 with Velocity Resolved Spectroscopy}

\author[0000-0003-1356-1096]{Elizabeth Tarantino}
\affiliation{Department of Astronomy, University of Maryland, College Park, MD 20742, USA}

\author[0000-0002-5480-5686]{Alberto D. Bolatto}
\affiliation{Department of Astronomy, University of Maryland, College Park, MD 20742, USA}

\author[0000-0002-2775-0595]{Rodrigo Herrera-Camus}
\affiliation{Astronomy Department, Universidad de Concepci\'{o}n, Barrio Universitario, Concepci\i{o}n, Chile}

\author[0000-0001-6159-9174]{Andrew I. Harris}
\affiliation{Department of Astronomy, University of Maryland, College Park, MD 20742, USA}

\author[0000-0003-0030-9510]{Mark Wolfire}
\affiliation{Department of Astronomy, University of Maryland, College Park, MD 20742, USA}

\author[0000-0002-2064-7691]{Christof Buchbender}
\affiliation{KOSMA, I. Physikalisches Institut, Universit\"{a}t zu K\"{o}ln, Z\"{u}lpicher Stra{\ss}e 77, D-50937 K\"{o}ln, Germany}

\author[0000-0002-5258-7224]{Kevin V. Croxall}
\affiliation{Expeed Software, Columbus, OH}

\author[0000-0002-5782-9093]{Daniel A. Dale}
\affiliation{Dept. of Physics and Astronomy, University of Wyoming, Laramie, WY, USA}

\author[0000-0002-9768-0246]{Brent Groves}
\affiliation{International Centre for Radio Astronomy Research, University of Western Australia, 7 Fairway, Crawley, 6009, WA, Australia}

\author[0000-0003-2508-2586]{Rebecca C. Levy}
\affiliation{Department of Astronomy, University of Maryland, College Park, MD 20742, USA}

\author[0000-0001-5389-0535]{Denise Riquelme}
\affiliation{Max-Planck-Institut für Radioastronomie, Auf dem H\"{u}gel 69, D-53121 Bonn, Germany}

\author[0000-0003-1545-5078]{J.-D T. Smith}
\affiliation{Dept. of Physics \& Astronomy, University of Toledo, Toledo, OH 43606, USA}

\author[0000-0001-7658-4397]{J\"{u}rgen Stutzki}
\affiliation{KOSMA, I. Physikalisches Institut, Universit\"{a}t zu K\"{o}ln, Z\"{u}lpicher Stra{\ss}e 77, D-50937 K\"{o}ln, Germany}

\correspondingauthor{Elizabeth Tarantino}
\email{ejtino@astro.umd.edu}

\begin{abstract}

\changes{The \cii\ fine-structure transition at 158 \micron\ is frequently the brightest far-infrared line in galaxies. Due to its low ionization potential, $\mathrm{C^+}$ can trace the ionized, atomic, and molecular phases of the ISM. We present velocity resolved \cii\ and \nii\ pointed observations from SOFIA/GREAT on $\sim$500~pc scales in the nearby galaxies M101 and NGC~6946 and investigate the multi-phase origin of \cii\ emission over a range of environments. We show that ionized gas makes a negligible contribution to the \cii\ emission in these positions using \nii\ observations. We spectrally decompose the \cii\ emission into components associated with the molecular and atomic phases using existing CO(2--1) and \hi\ data and show that a peak signal-to-noise ratio of 10--15 is necessary for a reliable decomposition. In general, we find that in our pointings $\gtrsim$50\% of the \cii\ emission arises from the atomic phase, with no strong dependence on star formation rate, metallicity, or galactocentric radius. We do find a difference between pointings in these two galaxies, where locations in NGC~6946 tend to have larger fractions of \cii\ emission associated with the molecular phase than in M101. We also find a weak but consistent trend for fainter \cii\ emission to exhibit a larger contribution from the atomic medium. We compute the thermal pressure of the cold neutral medium through the \cii\ cooling function and find $\log(P_{th}/k)=3.8-4.6\mathrm{~[K ~ cm^{-3}]}$, a value slightly higher than similar determinations, likely because our observations are biased towards star-forming regions.}

% suggesting that atomic gas associated with the \cii\ emission in our measurements, which are weighted towards the star forming regions, is at a slightly higher density and pressure than the average cold neutral medium in similar galaxies

\end{abstract}


\section{Introduction}

\label{sec:intro}
Emission from the far-infrared (FIR) \cii\ 158 \micron\ line is bright and ubiquitous in most star-forming galaxies. It is the $^2P_{3/2}^0 \rightarrow^2P_{1/2}^0 $ fine-structure, collisionally excited line of singly ionized carbon, $\mathrm{C^+}$. The emission from \cii\ provides a major cooling channel for the gas in the interstellar medium (ISM), specifically in the cold neutral medium \citep{Wolfire2003}, on the illuminated surfaces of molecular clouds, and, along with [OI], in dense photodissociation regions (PDRs) \citep{Tielens1985, hollenbach1999}. \cii\ is often the brightest emission line in the FIR from galaxies, amounting to about 0.1\% - 1\% of the integrated FIR continuum emission \citep{ crawford1985, Stacey1991}. Previous studies have also shown a correlation between \cii\ emission strength and star formation rates \citep{Stacey1991,Boselli2002,DeLooze2014, Herrera2015, Herrera2018, Smith2017}. In low metallicity environments, the \cii\ line is the only coolant necessary to form stars \citep{Glover2012, Krumholz2012}. Studying the nature of \cii\ emission is thus vital to the understanding of star formation and cooling in the ISM.

Ionized carbon ($\mathrm{C^+}$) can be present throughout the different phases of the ISM due to the low ionization potential of neutral carbon (11.26 eV, slightly less than that of hydrogen). Collisions with electrons (e$^-$), neutral hydrogen (\hi), and molecular hydrogen (\htwo) produce \cii\ emission that is found in the warm ionized medium, the neutral atomic medium, and cold molecular gas, respectively \citep[e.g.,][]{Madden1993, Heiles1994, Kim2002,pineda2013}. Identifying the contribution that each of the phases have to the overall \cii\ intensity allows one to determine which phase is dominant. Such information can be used to determine the physical conditions of the ISM, such as the thermal pressure of the constituent phases \citep[e.g.,][]{goldsmith2012, pineda2013, cormier2015, cormier2019, Lebouteiller2019, Sutter2019}.

The \cii\ emission is partially associated with the tracer of molecular gas, CO, as seen in correlations between the intensity of \cii\ and CO \citep{Wolfire1989, Stacey1991, accurso2017, Zanella2018}. On the surfaces of molecular clouds, far ultra-violet (FUV) radiation fields can dissociate CO into C and O and photoionize C to $\mathrm{C^+}$ while the hydrogen remains in molecular (\htwo) form, producing \cii\ emission associated with the CO-traced molecular cloud. These PDRs are very bright and will account for most of the \cii\ emission close to massive star formation \citep[e.g.,][]{Tielens1985}. \citet{pineda2013, Pineda2014} use velocity resolved \textit{Herschel}/HIFI \cii\ observations in the plane of the Milky Way to quantify the degree to which \cii\ is associated with the CO and find that 30-47\% of the total \cii\ emission observed comes from the molecular gas near dense PDRs. The study by \citet{deBlok2016} compared \textit{Herschel}/PACS \cii\ observations of 10 galaxies to CO and \hi\ data and found that the \cii\ radial surface density profiles are shallower than CO but much steeper than the \hi\ surface density profile. At low metallicities, however, the \cii\ associated to the CO-emitting molecular gas can be more complex. A decrease in the dust abundance leads to less shielding of CO clouds, creating regions of molecular material that produce \cii\ emission but faint in CO, called ``CO-dark" or ``CO-faint" gas \citep{grenier2005, Wolfire2010, Jameson2018, Madden2020}. For example, analysis of velocity resolved \cii\ observations in the low metallicity dwarf galaxy NGC~4214 suggests that 79\% of the molecular mass is traced by \cii\ alone, whereas only 21\% is traced by CO \citep{fahrion2017}.

Ionized carbon fine-structure emission also arises from the atomic medium and plays an important role in the radiative heating and cooling balance of this phase \citep{Wolfire2003}. The atomic gas, as traced by the hyperfine 21 cm spin flip HI transition, has a dense cold component (Cold Neutral Medium, CNM; $n_{\rm{H}}\approx 50$ cm$^{-3}$, $T_{\rm{kin}}\approx 80$ K) and a diffuse warm component (Warm Neutral Medium, WNM; $n_{\rm{H}}\approx 0.5$ cm$^{-3}$, $T_{\rm{kin}}\approx 8000$ K) in approximate pressure equilibrium with one another \citep{field1969, wolfire1995, heiles2003}. Because of the difference in volume densities, the contribution from the WNM to the overall \cii\ emission is $\sim$20 times less than that of the CNM \citep{Wolfire2010, pineda2013, fahrion2017, Lebouteiller2019}. Thus the \cii\ line can be used to directly probe the conditions of the CNM. Once the CNM is isolated, we can estimate the thermal pressure, which is related to the star formation rate, metallicity, and the thermal balance between the heating and cooling of the system \citep{Wolfire2003, ostriker2010}. In observations of star forming regions with spatial resolutions of a few parsecs, the atomic gas contributes 5\% - 15\% to the overall \cii\ emission \citep{okada2015, okada2019, requena2016, Lebouteiller2019}. In contrast, \cii\ observations of regions that are more quiescent or at larger resolutions of 50-200 parsecs, associate 20\% - 46\% of the \cii\ emission to the atomic phase \citep{Kramer2013, pineda2013, fahrion2017}. The spatial resolution and star formation activity may therefore play a role when decomposing the \cii\ emission. 

 \begin{figure*}[t!]
\includegraphics[width=1\textwidth]{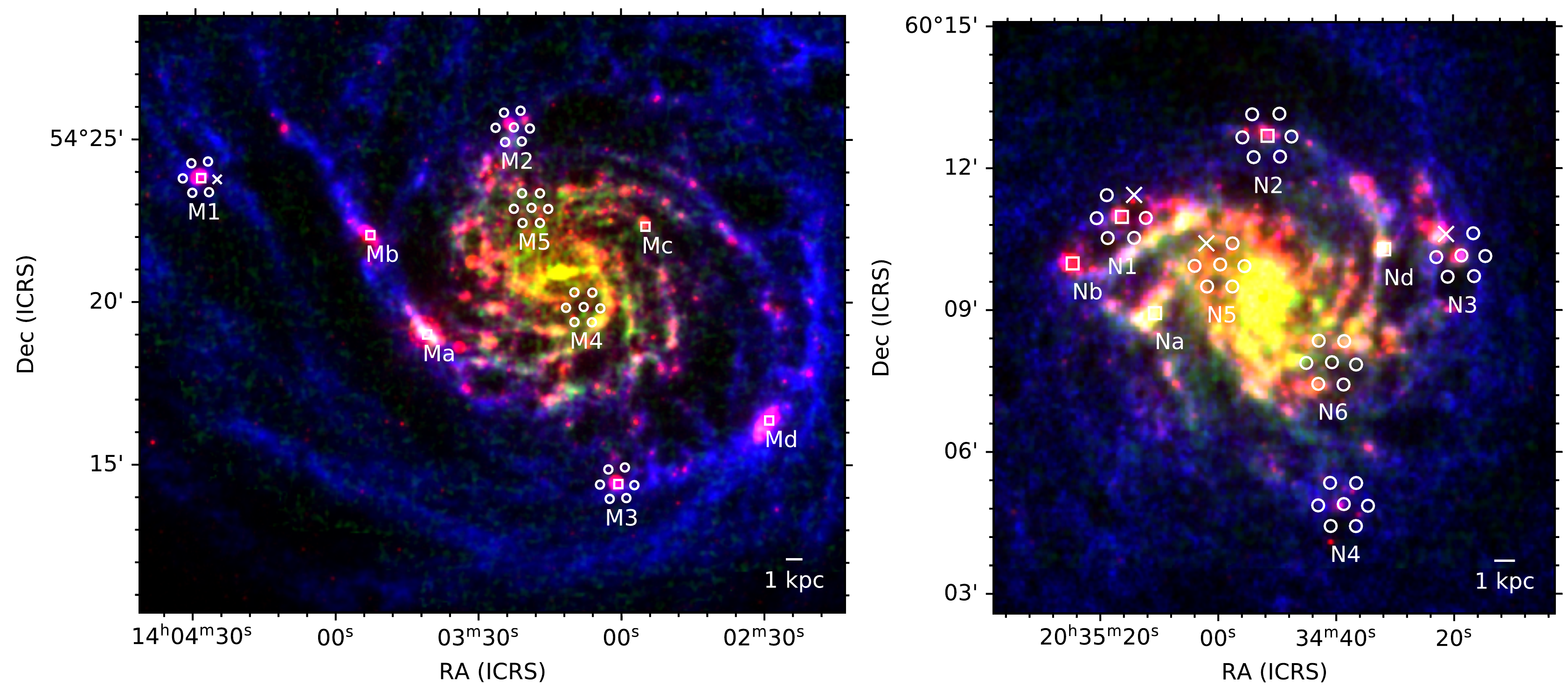}
\caption{M101 (left) and NGC 6946 (right) three-color image with the SOFIA/GREAT pointings overlaid. The three-colors show the ancillary datasets used in this work: the 24 \micron\ map from \textit{Spitzer} (red), the \hi\ column density map from the THINGS survey (blue), and the CO moment 0 map from the HERACLES survey (green). SOFIA cycle-4 observations (7 pixel upGREAT pointings) are labeled with numbers while the cycle 2 observations (single pointing with GREAT) are labeled with letters. \changes{Region Nc is not shown because the GREAT receiver was mistuned at that position and the region was re-observed as N1 in cycle 4.} Pointing circles are the approximate beam of these data and correspond to the quiescent regions (\changes{\sSFR\ $< 6\times10^{-2}$} \msun\ yr$^{-1}$ kpc$^{-2}$), squares represent the star-forming regions (\changes{\sSFR\ $> 6\times10^{-2}$} \msun\ yr$^{-1}$ kpc$^{-2}$), and x are regions that were removed from the sample due to emission in the off-source chop-position contaminating the observed profile. Pointing selection shows a range of environments, probing the metallicity gradient in each galaxy and different levels of star formation rate.}
\label{fig:gal}
\end{figure*}

The contribution the ionized gas has to the \cii\ emission is usually found through observations of the \nii\ line. Ionized nitrogen is only present in the ionized gas and its 205 \micron\ transition has a critical density similar to \cii, enabling \nii\ 205 \micron\ observations to isolate the contribution of ionized gas to the \cii\ emission \citep{Oberst2006}. \citet{Croxall2017} use observations of the \nii\ and \cii\ lines in a variety of galaxies to find that ionized gas contributes only $\sim$26\% to the \cii\ emission on average. Expanding on the sample of galaxies used by \citet{Croxall2017}, \citet{Sutter2019} identify that $\sim$33\% of the \cii\ emission comes from ionized gas by comparing the \nii\ and \cii\ measurements in these galaxies. Both \citet{Lebouteiller2019} and \citet{fahrion2017} find that the contribution ionized gas has to the \cii\ emission is negligible through a combination of \nii\ observations and modeling of the PDRs in each system. The ionized gas, however, makes a larger contribution of about 36\% - 75\% in the bright \hii\ region M17 SW and in the center of the starburst galaxy IC 342 \citep{perez2015, rollig2016}. Overall, it appears that the ionized gas tends to contribute a small amount to the \cii\ emission, except perhaps in areas of extended, dense ionized gas.

One method to identify the origin of the \cii\ emission is to compare the velocity profiles of the \cii , CO, \hi , and other tracers in order to quantify the contribution each phase has to the \cii\ emission. This method can use velocity resolved observations of \cii\ from the SOFIA/GREAT or \textit{Herschel}/HIFI instruments, along with similar resolution data for the tracers of component data, such as 21\,cm \hi\ for the atomic gas and CO for the molecular gas. Previous studies use this method but mostly target individual star-forming regions in the Milky Way \citep{perez2015}, Magellanic Clouds \citep{okada2015, okada2019, Pineda2017, Lebouteiller2019}, or the bright centers of nearby galaxies \citep{Mookerjea2016, rollig2016, fahrion2017}. There are few studies, however, that explore the origin of \cii\ in both star-forming and quiescent regions, an important regime due to the multi-phase nature of \cii. This work aims to spectrally decompose the \cii\ emission in a variety of environments, including different star formation rate surface densities (\sSFR) and metallicities, in the two galaxies M101 (NGC 5457) and NGC 6946 at a resolution of $\sim$500~pc. These galaxies are representative of spiral galaxies as a whole, are at distances of $6.8$~Mpc for M101 \citep{Fernandez2018} and $7.8$~Mpc for NGC 6946 \citep{Anand2018, Murphy2018}, and have a wealth of ancillary data available. We will also present an evaluation of the spectral profile decomposition technique.
 
The organization of this paper is as follows. In Section \ref{sec:observations} we describe the data used in the decomposition. Section \ref{sec:methodology} we discuss the method used for the decomposition and evaluate its accuracy. Section \ref{sec:decomp_all} shows the results of the decomposition and describes limits on the contribution of the ionized gas to the \cii\ emission. Section \ref{sec:discussion} computes the pressure of the CNM through the \cii\ cooling function and compares the results of the decomposition to other works. Lastly, Section \ref{sec:conclusions} summarizes our conclusions.

\section{Observations} \label{sec:observations}

\subsection{SOFIA Data}
Observations of M101 and NGC 6946 were taken using the German REceiver for Astronomy at Terahertz Frequencies (GREAT and its improved successor upGREAT) on board the Stratospheric Observatory for Infrared Astronomy (SOFIA) in cycles 2 and 4 \citep{Heyminck2012, Risacher2016, Risacher2018}. The cycle 2 (PI: Herrera-Camus, project 02\_0098) data used the original GREAT instrument, consisting of a single element receiver, and targeted four regions in M101 and four in NGC~6946. Observations were taken May 20th and 21st of 2014, with receivers tuned to 1900.5 GHz (\cii\ 158 $\mu$m) and 1461.1 GHz (\nii\ 205 $\mu$m). The receiver was accidentally mis-tuned for the [CII] observations in one of the regions (labeled Nc) and was then re-observed with upGREAT in cycle 4 \changes{as region N1}. We adopt a uniform beam size of 15\arcsec\ for \cii\ and 18\arcsec\ for \nii . The average sensitivity for \cii\ in cycle 2 was T$_{\rm mb}$ = 0.11 K and T$_{\rm mb} = 0.05 \ \rm K$ for \nii\ in a velocity channel width of 5.2 \kms . 
 
The cycle 4 data (PI: Bolatto, project 04\_0151) comprised of eleven total regions, five in M101 and six in NGC~6946, using the dual polarization upGREAT instrument on SOFIA. The upGREAT instrument consists of two seven-element hexagonal arrays, one for each polarization, and the two polarizations were averaged together for these data. The observations of M101 were taken during the upGREAT commissioning on December 9th and 10th, 2015 and the NGC 6946 observations were taken on May 12th, 18th, 19th, and 25th of 2016. For both galaxies, the Low Frequency Array (LFA) band was tuned to 1900.5 GHz (\cii\ 158 $\mu$m) and the L1 band was tuned to 1461.1 GHz (\nii\ 205 $\mu$m). The half-power beam widths were 15\arcsec\ for \cii\ and 18\arcsec\ for \nii. The average sensitivity for \cii\ achieved with the integration time obtained for each observation in cycle 4 is T$_{\rm mb}$ = 0.05 K and T$_{\rm mb}$ = 0.04 K for \nii\ in a velocity channel width of 5.2 \kms . Cycle 4 pointings are referred to by their region name and a number denoting the pointing number in the array (e.g. N2-0 would be the cycle 4 NGC 6946 region N2 and the 0th, central pointing). 

The single point chopped mode for SOFIA/GREAT was used for each cycle. We excluded pointings where the chop-off position showed weak emission, contaminating the on-source spectrum. The data in both cycles were processed with the eXtended bandwidth Fast Fourier Transform Spectrometer (XFFTS) and calibrated with the standard GREAT calibrator \citep{Guan2012}. The antenna efficiency for both cycles is $\eta_f = 0.97$ and the main beam efficiency varies between $\eta_{mb} = 0.65 - 0.71$, for the different elements in the array. The level 3 data products were produced through the CLASS/GILDAS software where a first order polynomial spectral baseline was removed.

\subsubsection{Pointing selection}

The placement of these pointings is shown in \autoref{fig:gal}. There are two types of regimes targeted in this study: areas that are coincident with high star formation rates (represented by squares in \autoref{fig:gal}) and the more quiescent regions found in the interarm of each galaxy (represented by circles in \autoref{fig:gal}). The star-forming regions are chosen for having strong star formation activity as traced by H$\alpha$, far-UV, and 24 $\mathrm{\mu}$m emission. The diffuse ISM in these regions is exposed to about six times higher average radiation fields than the solar neighborhood \citep{Draine2007, Aniano2020} when measured on large ($\sim$kpc) scales. The other category targeted are the quiescent interarm regions. These regions are selected for their low star formation activity, weak CO and HI emission, and have similar average radiation field strengths as the solar neighborhood \citep{Draine2007, Aniano2020}. \changes{We use a threshold value of \SFRcut , which roughly bisects the sample, to distinguish between the star-forming and quiescent regions.}

In addition to the star formation rate surface density (\sSFR), we are also interested in studying the effect of metallicity on the \cii\ decomposition. We use the abundance gradients published by \cite{Pilyugin2014}, which employs the ``strong-line method" of abundance determination on 130 nearby galaxies (including M101 and NGC 6946) to produce homogeneous gas phase oxygen abundance gradients from the \hii\ regions in these galaxies. Although there are small azimuthal metallicity variation in similar spiral galaxies, the overwhelming metallicity change is due to these radial gradients \citep{Kreckel2020}. \changes{Similar to the \sSFR , we use a threshold value of $\rm 12 + log(O/H) = 8.55$ to distinguish between high and low metallicities.} We note that the pointings centered on star-forming regions extend to larger galactocentric distance and are consequently biased towards lower metallicities. \changes{We cannot separate the effect of the lower metallicity on the high star formation regions because the lack of integration time on quiescent pointings at low metallicity leads to little \cii\ detections in this regime.}
% the integration time for 

\subsection{Herschel PACS [CII] data}
\label{sec:PACS}
\cii\ observations of M101 and NGC 6946 were also made with the \textit{Herschel}/PACS instrument as part of the KINGFISH program \citep{Kennicutt2011}. The KINGFISH program focused on deep spectroscopic imaging of ISM diagnostic lines, including \cii, at a resolution of $\sim$12\arcsec.

With an effective spectral resolution of about 220 \kms, PACS does not resolve the \cii\ line, in contrast to the GREAT instrument. However, we can compare the integrated intensities of the \cii\ line between the two instruments. The overlap between the \textit{Herschel}/PACS observations and detected SOFIA/GREAT \cii\ pointings includes all but three GREAT pointings. 

\begin{figure}
\centering
\includegraphics[width=\columnwidth]{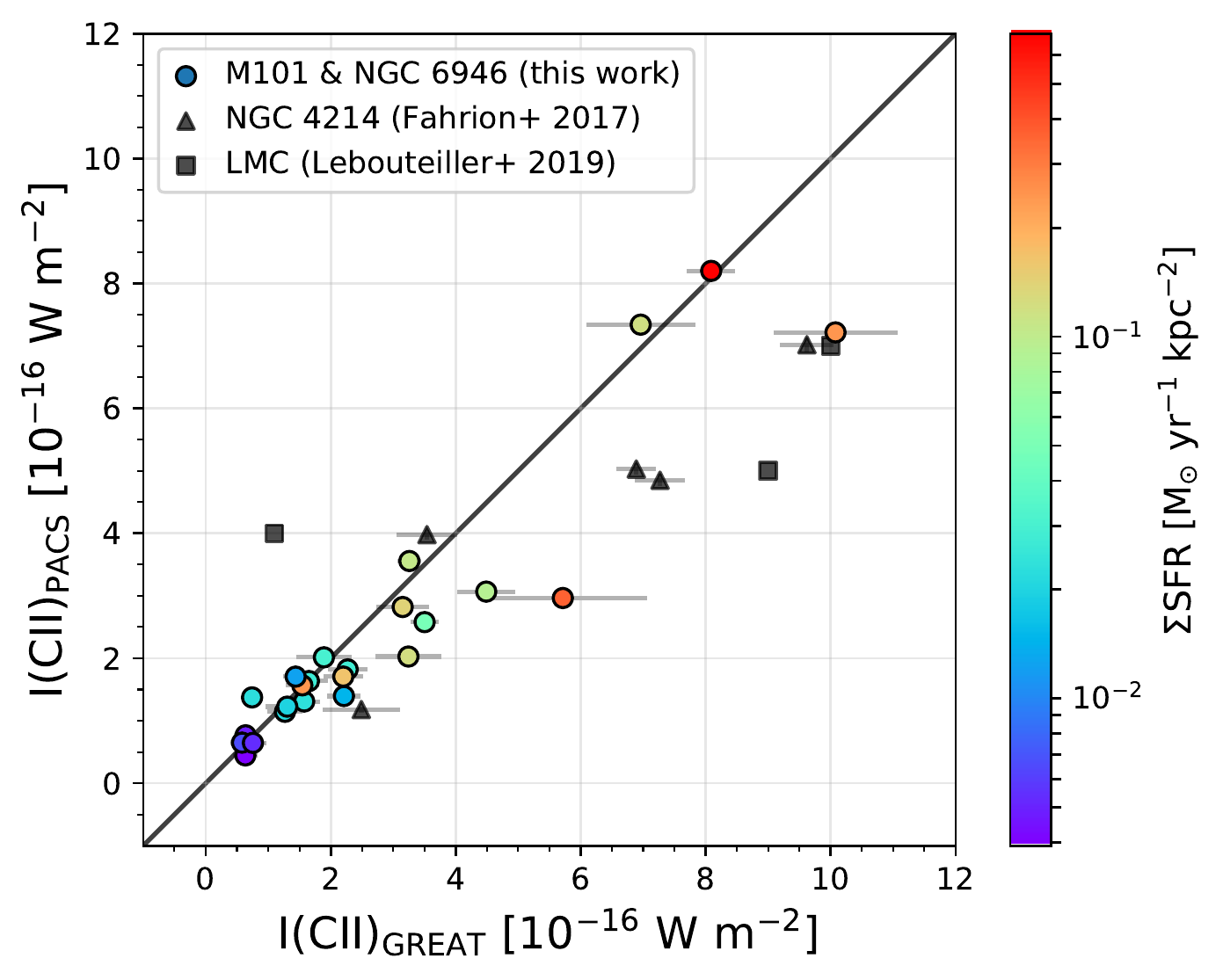}
\caption{The line flux of the SOFIA/GREAT \cii\ emission compared to the \textit{Herschel} PACS \cii\ flux. Circles are the data in this paper, triangles are from NGC 4214 \citep{fahrion2017}, and squares are for the LMC \citep{Lebouteiller2019}. The black line represents the line of unity. Colorscale is the \sSFR\ calculated from 24\micron\ data. The GREAT data in this work is $\sim$7\% brighter than the PACS data on average, but this difference is within absolute flux accuracy bounds for the PACS instrument \citep[15\%,][]{Croxall2013}. The GREAT and PACS \cii\ intensities from the literature have a larger discrepancy, likely due to approximating the PACS PSF as a Gaussian. Error bars reported are the statistical 1$\sigma$ rms noise (PACS errorbars are smaller than the data points). }
\label{fig:PACS}
\end{figure}

To compare the GREAT and PACS \cii\ intensities we use a convolution kernel to convert from the PACS 158\,\micron\ Point Spread Function (PSF) to the Gaussian 15\arcsec\ beam of the GREAT data from the kernels provided in \citet{Aniano2011}. These kernels are most appropriate for the PACS continuum camera, but are likely to result in a much better approximation than assuming a Gaussian PSF. \autoref{fig:PACS} shows the comparison between GREAT and PACS line integrated intensities with the black line representing the line of unity. We also show data from the literature for NGC~4214 and the Large Magellanic Cloud (LMC) for comparison \citep{fahrion2017,Lebouteiller2019}. Error bars are reported where available and correspond to the 1$\sigma$ rms noise of the given spectrum. 

There is a small systematic difference of about 7\% between the GREAT flux and the corresponding PACS \cii\ flux. The expected absolute flux uncertainties for the PACS KINGFISH data are about 15\%, making this discrepancy within the bounds of the PACS and GREAT calibration uncertainty \citep{Croxall2013}. The observations from the literature, by contrast, show a much larger discrepancy, with GREAT data approximately 40\% brighter than PACS. These studies, however, approximate the PACS PSF as a Gaussian beam instead of using the more accurate convolution kernels required to transform the PACS PSF into a Gaussian comparable to the GREAT beam. The native PACS spectrograph PSF at 158\,\micron\ is not Gaussian shaped: a significant portion of the total power resides in wide wings and a high pedestal that creates a halo around a point source \citep{Geis2010}. When we do not use the proper PACs convolution kernel, we receive a similar discrepancy of $\sim40\%$ that is seen in the observations from the literature. Therefore, we attribute their discrepancy to a beam mismatch.  

\begin{figure*}
\centering
\includegraphics[width=\textwidth]{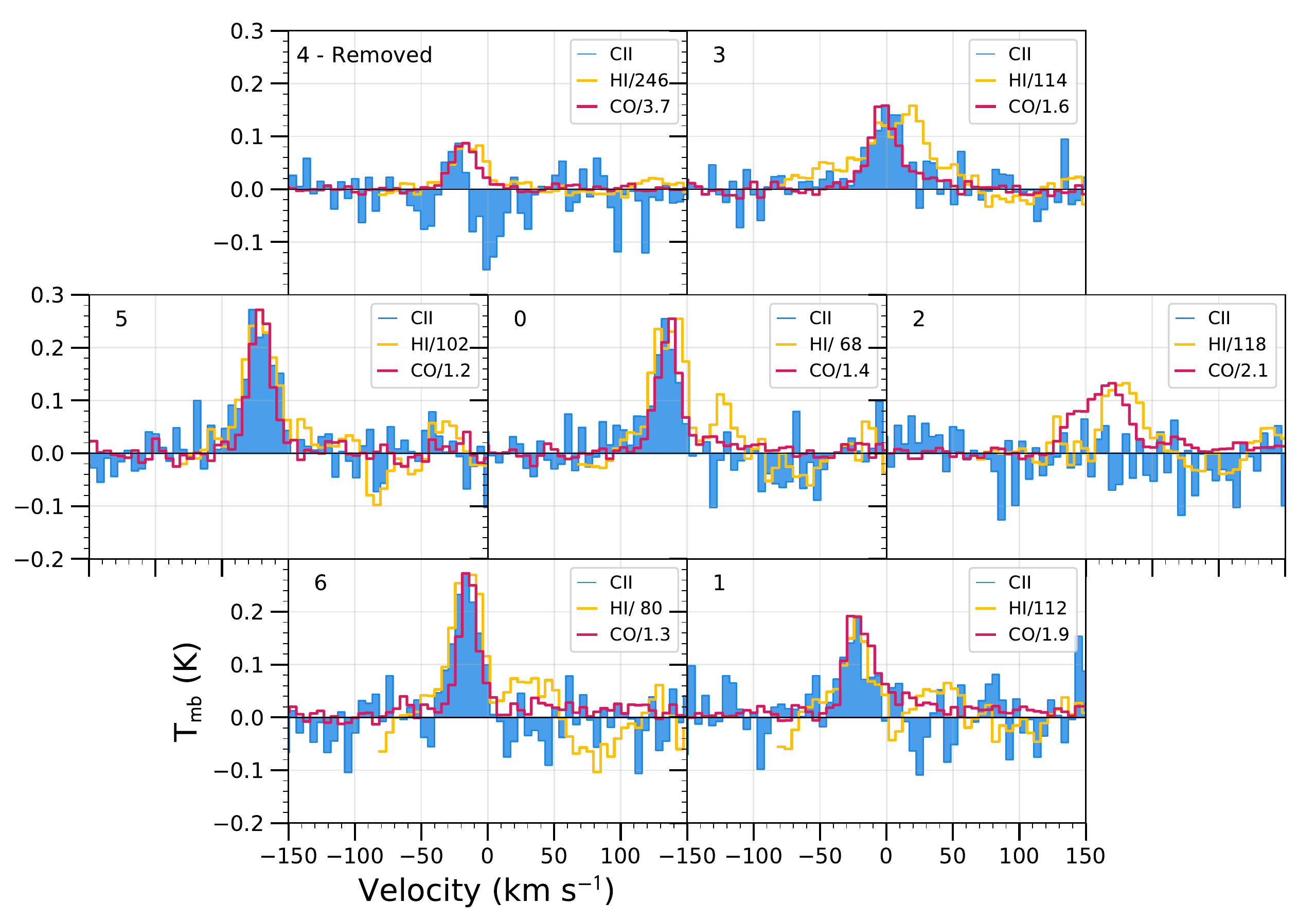}
\caption{An example of spectra from the upGREAT receiver in NGC 6946 region N5. This region contains detected \cii\ emission in most of the pointings. The HI and CO spectra are normalized to the maximum intensity of the \cii\ spectra. The width of the \cii\ profiles tend to lie in between the CO and HI profile widths, suggesting an origin from both the atomic and molecular gas. These offsets between the \cii , CO, and \hi , are greater than the instrument's spectral resolution of 5.2 \kms . Pointing number is labeled in the top left corner of each spectrum. Region N5-4 contains negative emission from emission in the off during calibration and is an example of a spectrum that is removed from this analysis.}
\label{fig:ex_spec}
\end{figure*}

\subsection{\hi\ 21 cm data}
The \hi\ data comes from The \hi\ Nearby Galaxy Survey (THINGS, \citealt{Walter2008}), a 21 cm (1.4204 GHz) \changes{Very Large Array (VLA)} survey of nearby galaxies that uses the same galaxy sample as SINGS, the Spitzer Infrared Nearby Galaxies Survey \citep[]{Kennicutt2003}. We use the THINGS naturally weighted cubes which have a half-power beam width of $10.8\arcsec \times 10.2\arcsec$ for M101 and $6.0\arcsec\times 5.6\arcsec$ for NGC~6946. We will use the HI data as the tracer for the atomic material in these galaxies.

Maps of extended objects made with interferometer data that are not combined with single dish data have missing flux on large scales, called the short spacings problem \citep{braun1985}. This may manifest as a shallow negative bowl around the emission, caused by the interferometer filtering out the lowest spatial frequencies. The effect is seen in the \hi\ THINGS data of these galaxies and is particularly strong in NGC 6946. We cannot properly correct for the lack of this information, but we can mitigate the effect of the negative portions of the spectrum in our analysis. We fit the negative spectral region around the signal with a first order polynomial and add the resulting fit to the negative wings that border the signal. For our analysis, which focuses on the \changes{spectral} shape of emission, this fairly small re-baselining correction is sufficient to avoid negative regions having a strong effect on the decomposition.

\subsection{CO (2-1) data}
The CO data are from the HERA CO-Line Extragalactic Survey (HERACLES, \citealt{Leroy2009}). This is a CO(2-1) (230.54 GHz) survey using the IRAM 30~m telescope, designed to complement the THINGS and SINGS surveys. The half-power beam width for both galaxies is 13$\arcsec$. The CO data will be used as the tracer for the molecular material in these galaxies. 

\subsection{Star formation Rates}
\changes{We use a combination of 24\,\micron\ and H$\alpha$ data in order to trace the obscured and unobscured star formation activity in these galaxies. The 24\,\micron\ data were taken from the SINGS Survey \citep[]{Kennicutt2003} and we use the convolution kernels provided by \citet[]{Aniano2011} to convert the MIPS 24\,\micron\ PSF into a Gaussian beam of 15\arcsec . The H$\alpha$ data were compiled and processed by \citet[]{Leroy2012}, where the map for NGC~6946 came from the SINGS Survey \citep[]{Kennicutt2003} and the map for M101 was retrieved from \citet[]{Hoopes2001}. The contribution of \nii\ to the H$\alpha$ emission was removed \citep[]{Kennicutt2008, Kennicutt2009}, the foreground stars were subtracted \citep[]{Munoz2009}, and the H$\alpha$ data were corrected for Galactic extinction \citep[]{Schlegel1998}. We use the calibration from \citet[]{calzetti2007} (Equation~7) and the combination of the 24\,\micron\ and H$\alpha$ data to calculate the star formation rate surface densities (\sSFR) in these galaxies. The \sSFR\ is corrected for inclination, assuming a value of 38$^{\circ}$ for NGC~6946 and 18$^{\circ}$ for M101.}

% and ancillary H$\alpha$ data to estimate the star formation rate surface density (\sSFR) for these regions. We use the convolution kernels provided by \citet[]{Aniano2011} to convert the MIPS 24\,\micron\ PSF into a Gaussian beam of 15\arcsec, then used calibrations in \cite{calzetti2007} to convert the 24\,\micron\ luminosity per unit area, $\Sigma_{24}$ in erg\,s$^{-1}$\,kpc$^{-2}$, into a \sSFR\ in \msun\,yr$^{-1}$\,kpc$^{-2}$:

% \begin{equation}
% \sSFR = 1.56 \rm \times 10^{-35} \,\Sigma_{24}^{0.8850}.
% \end{equation}

\changes{This calibration adopts a truncated Salpeter IMF with a slope of 1.3 in the range of 0.1–0.5~\msun\ and a slope of 2.3 in the range of 0.5–120~\msun .} The distribution of \sSFR\ is shown in the colorscale of \autoref{fig:PACS}, where there is a spread of about 3 orders of magnitude in \sSFR\ for our sample. 

\subsection{Matching the spectral and spatial resolution}
In order to match the resolution of the SOFIA data, we convolved the CO and \hi\ maps with a Gaussian to the GREAT beam size of 15\arcsec, which corresponds to 495~pc for M101 and 567~pc for NGC~6946. 

We also resample the \cii , CO, and \hi\ spectra to a common velocity resolution of 5.2\,\kms\ by hanning smoothing (when applicable) and regrid each spectra to match that of the 5.2\,\kms\ resolution CO data. All data were in the radio velocity convention and when necessary we converted the velocity reference to the kinematic local standard of rest (LSRK). An example of the three spectra after smoothing to the same resolution is shown in \autoref{fig:ex_spec}.

\subsection{Selecting the spectra for this study}
\label{sec:sample}
We select the regions for the analysis by the integrated intensity of the \cii\ line. First, we remove all spectra that exhibit features due to emission in the off position (see \autoref{fig:ex_spec}, region N5-4) or spectra with noise spikes greater than 1\,K. We calculate the \cii\ integrated intensity and 1$\sigma$ rms error by defining the bounds of integration from the \hi\ data. We then select the spectra that have an integrated intensity greater than three times the calculated rms 1$\sigma$ noise level as the main sample in this work (referred to as the 3$\sigma$ sample). In addition, we use an integrated intensity cut, which includes all spectra greater than a given K \kms\ value, and a sample which uses all of the spectra, to test how the sample selection alters the results (see more in \S\ref{sec:stack}).

A summary of all the \cii\ spectra, including the position of pointings, star formation rate surface density, and the \cii\ integrated intensity are given in \autoref{tab:sum_tab}. 

\startlongtable 
\begin{deluxetable*}{ccccccccc} 
\tabletypesize 
\footnotesize 
\tablecaption{\cii\ SOFIA/GREAT Spectra Summary \label{tab:sum_tab} } 
\tablehead{ 
\colhead{Galaxy} & \colhead{Region} & \colhead{Cycle} & \colhead{R.A.} & \colhead{Decl.} & \colhead{12+log(O/H)} & \colhead{$\mathrm{\Sigma_{SFR}}$} & \colhead{$\mathrm{\int}$I$_\mathrm{[CII]}$} & \colhead{rms}  
\\ & & & \colhead{(J2000)} & \colhead{(J2000)} & & \colhead{(\msun\ yr$^{-1}$ kpc$^{-2}$)} & \colhead{(K \kms)}  & \colhead{(K)}   }
\startdata 
M101&Ma&2&14h03m41.0s&54d19m01.0s&8.43&6.88 $\rm \times \ 10^{-1}$&27.5 $\pm$ 1.3&0.06 \\
M101&M3-0&4&14h03m00.8s&54d14m25.2s&8.31&1.06$\rm \times \ 10^{-1}$& 7.5 $\pm$ 1.1&0.05 \\
NGC6946&Nd&2&20h34m32.0s&60d10m16.0s&8.53&1.18 $\rm \times \ 10^{-1}$&23.6 $\pm$ 3.0&0.12 \\
NGC6946&N5-5&4&20h35m04.3s&60d09m55.1s&8.62&4.54 $\rm \times \ 10^{-2}$& 8.0 $\pm$ 0.8&0.04 \\
NGC6946&N1-4&4&20h35m19.2s&60d11m24.7s&8.48&4.93 $\rm \times \ 10^{-3}$&-0.3 $\pm$ 1.2&0.06 \\
\enddata 
\tablecomments{Description of [CII] spectra with examples given from the two different cycles and galaxies used in this work. Region N1-4 represents a ``non-detected'' [CII] spectra due to the negative integrated [CII] intensity. Divide by 1.43 $\rm \times \ 10^{5}$ to convert $\mathrm{\int}$I$_\mathrm{[CII]}$ from K \kms\ to erg s$^{-1}$ cm$^{-2}$ sr$^{-1}$ \citep{goldsmith2012}. (This table in its entirety is available in a machine-readable form online.)}
\end{deluxetable*}

\section{Methodology} \label{sec:methodology}

\subsection{\cii\ emission decomposition description}
\label{sec:decomp}

The method of using the kinematic information to establish the origin of the \cii\ is presented in several analyses \citep[e.g.,][]{okada2015,okada2019,fahrion2017,Lebouteiller2019}. These approaches generally rely on decomposing the profiles into Gaussian components that can then be related with the \hi\ or CO spectra. Here we present another approach, by creating a model \cii\ spectrum that is comprised of a linear combination of the CO and \hi\ spectra, and finding the coefficients that best reproduce the \cii\ spectrum, similar to the work by \citet{Mookerjea2016}. This has the advantage of being entirely non-parametric, and of presenting a mathematically well-posed problem with a unique solution that lends itself to a simple reliability analysis. The drawback is that components that are not represented in our model (besides the \hi\ or the CO spectra) are not easily analyzed. To account for this, we show that the ionized gas has a negligible contribution to the \cii\ emission in \S\ref{sec:ionizedgas}. Because \hi\ has two phases that contribute equally to the 21\,cm spectrum, but \cii\ emission is thought to be predominately associated with one of them (the CNM), this method requires that we work on scales ($\sim$~500 pc) that are large enough for the phases to be well-mixed so that the kinematics of the 21\,cm emission represents well the CNM. 

We use the Rayleigh-Jeans brightness temperatures as a measure of the flux for the \cii , CO, and \hi\ spectra. The decomposition creates a model \cii\ spectrum from a linear combination of the CO and \hi\ spectra, where $w_{\rm CO}$ and $w_{\rm HI}$ are the constants for the linear combination:

\begin{equation}
\label{eq:comp}
    T_{\rm{[CII]}, \rm{model}} = w_{\rm{CO}}T_{\rm{CO}} + w_{\rm{HI}}T_{\rm{HI}}.
\end{equation}
We define $T_{\rm{CO}}$ and $T_{\rm{HI}}$ as the Rayleigh-Jeans brightness temperatures of the CO and \hi\ data, respectively, and $T_{\rm{[CII], model}}$ is the model \cii\ spectrum. We then use $\chi^2$ minimization to estimate the values of $w_{\rm{CO}}$ and $w_{\rm{HI}}$ that best reproduce the observed \cii\ spectrum given the noise of the observations:

\begin{equation}
    \chi^2 = \sum_{n=1}^{n} \frac{ (T_{\rm{[CII]}} - w_{\rm{CO}}T_{\rm{CO}} - w_{\rm{HI}}T_{\rm{HI}})^2}{\sigma_{\rm{[CII]}}^2 +  w_{\rm{CO}}^2\sigma_{\rm{CO}}^2 + w_{\rm{HI}}^2\sigma_{\rm{HI}}^2}
\label{eq:chi}
\end{equation}
where $\sigma$ corresponds to the rms noise of each spectrum and the model is evaluated across the $n$ channels in the given spectra. 

With best fit $w_{\rm{CO}}$ and $w_{\rm{HI}}$ values, we then calculate the fraction of the integrated \cii\ intensity associated with the molecular and atomic gas:

\begin{equation}
\label{eq:frac}
    f_{\rm{mol}} = \frac{w_{\rm{CO}} \int T_{\rm{CO}} dv}{\int T_{\rm{[CII]}}dv}; \;
    f_{\rm{atomic}} = \frac{w_{\rm{HI}} \int T_{\rm{HI}} dv}{\int T_{\rm{[CII]}}dv}.
\end{equation}

By using the linear combination of CO and \hi\ spectra as the model for the \cii\ spectra, we maximize the contribution the CO-traced molecular gas and the \hi -traced atomic gas have to the overall \cii\ emission. Additional contributing ISM components to the \cii\ emission will be seen in residuals of the fit if the velocity profiles \changes{have} a different shape from the CO or \hi\ profiles. ISM components that have similar velocity profiles as the CO or \hi\ will therefore be attributed to the tracer with the most similar shape. For example, \cii\ associated with the dense ionized gas from \hii\ regions will likely share a similar spectral profile to the CO that is associated with the dense PDRs. Thus, this dense ionized gas may be assigned to the molecular component.

\subsection{Evaluation of the decomposition method}\label{sec:eval}

\begin{figure}
\centering
\includegraphics[width=\columnwidth]{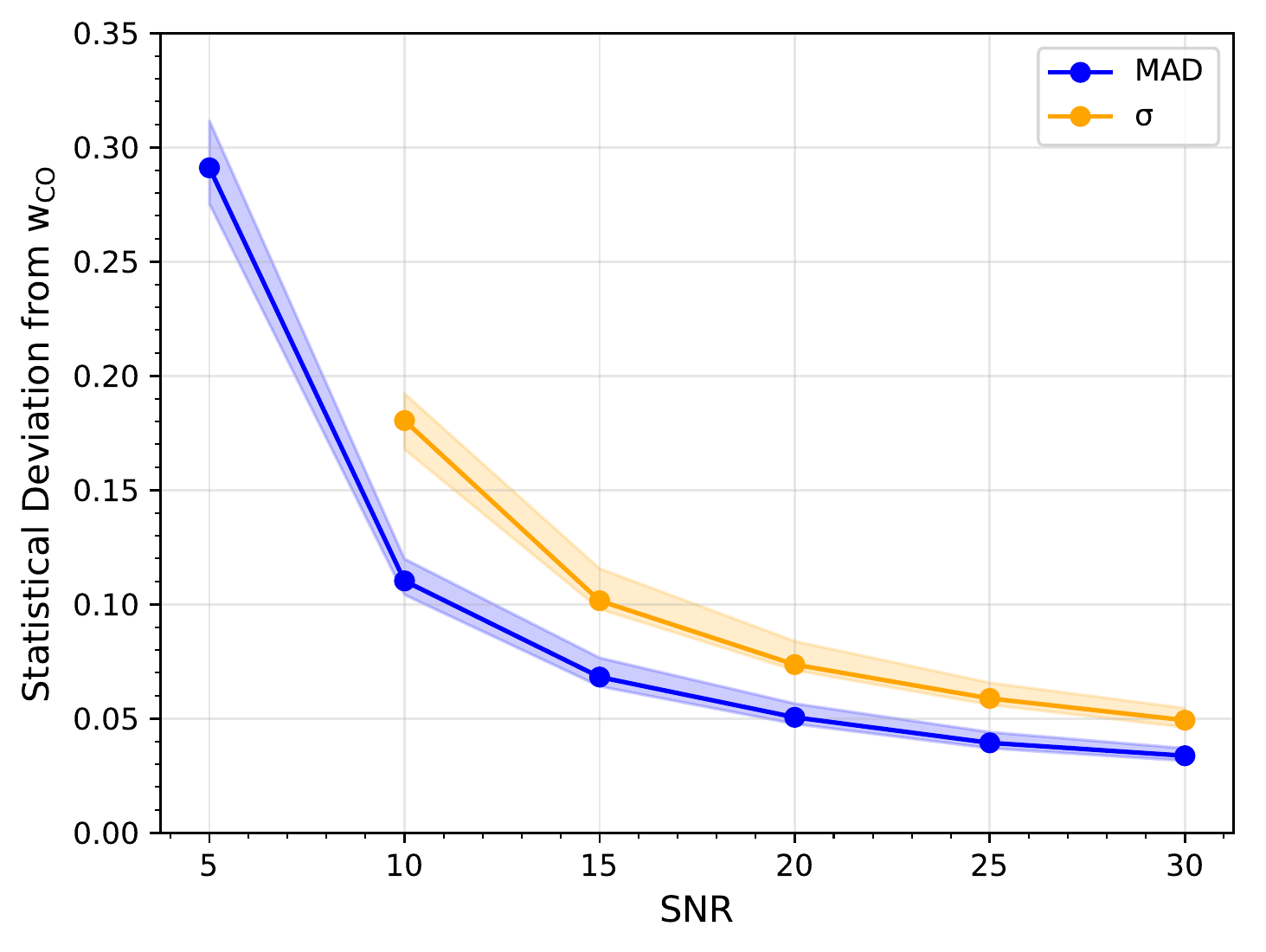}
\caption{Accuracy of the \cii\ decomposition method: error in the recovery of the input parameter in Monte Carlo realizations using realistic template spectra and varying SNR. The orange curve is the standard deviation of the fitted parameter with respect to the input. The blue curve is the median absolute deviation (MAD) of the fitted parameters. The shaded regions correspond to the variation of the statistical deviations from the $w_{CO}$ value. A SNR of $\sim15$ is necessary for recovering the fraction of emission from the molecular phase to an accuracy of $\Delta w_{CO}\approx\pm0.10$ at $1\sigma$.}\label{fig:accuracy}
\end{figure}

In order to evaluate the accuracy of the method, we run a series of simulated \cii\ decomposition cases to explore how this method changes with different parameters, such as the peak signal-to-noise ratio (SNR). We produce realistic CO and \hi\ templates by averaging the spectral profiles of our existing CO and \hi\ data, normalizing them to a peak of unity. Using a combination of these CO and \hi\ template spectra, we create simulated \cii\ spectra using different values of $\rm w_{CO}$ and $w\rm _{HI}$, where $w\rm{ _{CO} }+ w\rm{_{HI}} = 1$. We then add Gaussian distributed noise that correspond to the given \cii\ SNR for that trial. The input $w\rm _{CO}$ parameter ranges from 0.0~-~1.0 in 0.1 increments and the peak SNR ranges from 5 - 30 in increments of 5. Lastly, we use the $\chi^2$ minimization in \autoref{eq:chi} to calculate best fit values for $w_{\rm CO}$ and repeat the process 5000 times.

We find the accuracy of the decomposition method by comparing the input $w\rm _{CO}$ parameter to the resulting fitted parameter (note that this is a one-parameter problem since $w\rm _{CO} + w\rm _{HI} = 1$, so our results for the molecular fraction also apply to the atomic fraction). We calculate the standard deviation and median absolute deviation between the fitted $w\rm _{CO}$ and the respective input $w\rm _{CO}$. \autoref{fig:accuracy} shows both statistics averaged over all $w\rm _{CO}$ input parameters in a given SNR bin with the distribution of the $w\rm _{CO}$ input parameter represented by the shaded region. The standard deviation is more sensitive to outliers and consequently can have very large values, such as $\sigma$ = 660 for the SNR = 5 bin (not shown on figure). The median absolute deviation is not as sensitive to outliers \changes{and returns a value of 0.29 for SNR = 5 (equivalent to 0.41 for standard deviation when assuming Gaussian distributed data). The MAD value for SNR = 5, however, is much larger than for other SNR values, indicating that spectra with SNR = 5 do not give accurate results.}

\begin{figure}
\centering
\includegraphics[width=\columnwidth]{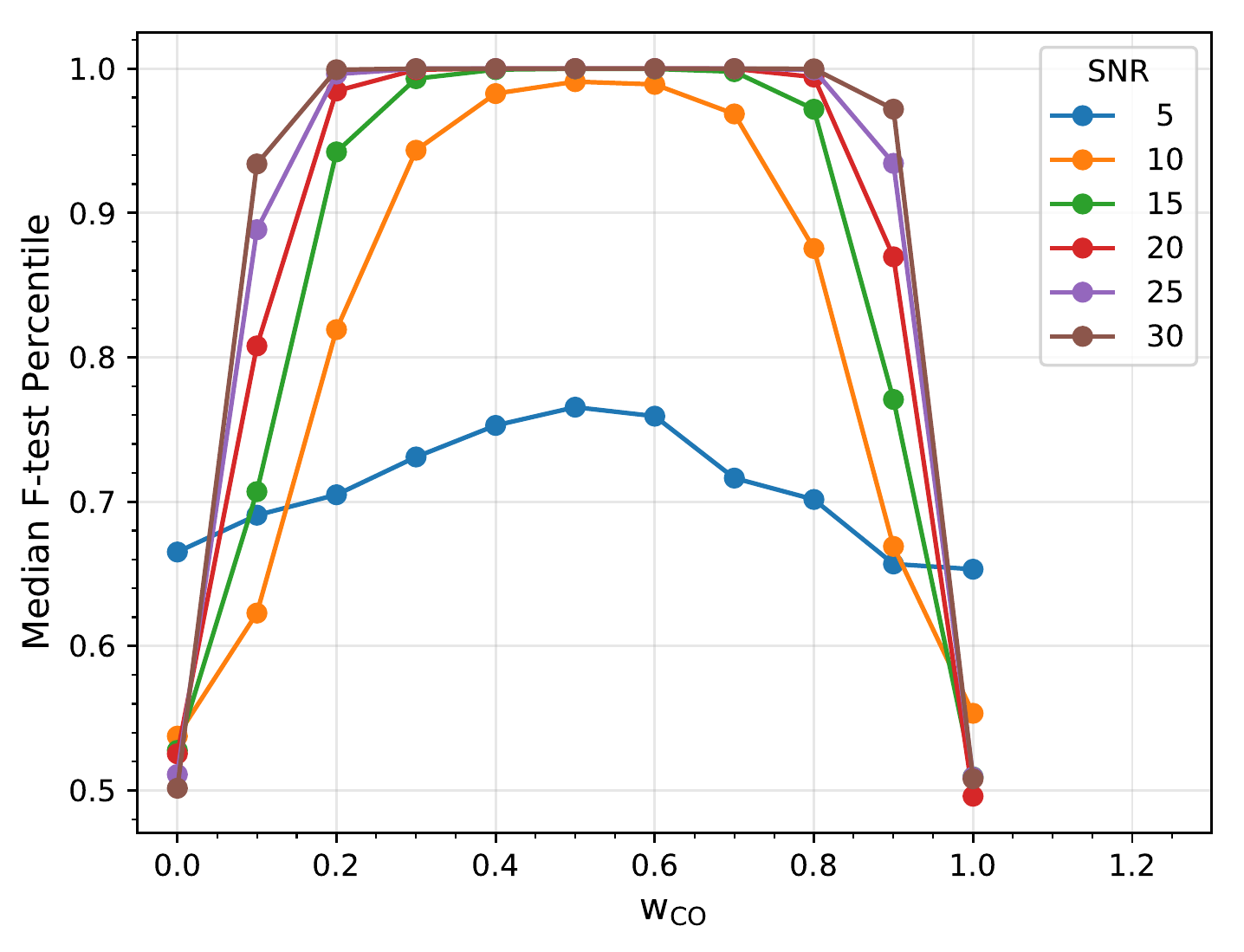}
\caption{When is a two-component description statistically better than a one-component model? A high F-test value shows that the two-component model provides a better description of the data. In turn, the ability to make this distinction requires a minimum SNR from the data. A two-component model is also more easily distinguishable from a single-component when both components have similar weights (note that $w_{\rm{CO}} + w_{\rm{HI}} = 1$). This plot shows that in cases where one-component contributes 20\% of the emission and the other 80\% a SNR$\sim$15 is necessary. For $40\%-60\%$ contributions this can be relaxed to SNR$\sim10$, but to distinguish between one and two components when the lesser component contributes only 10\% of the signal requires very high SNR$\gtrsim30$.}
\label{fig:unique}
\end{figure}

The standard deviation represents the 1$\sigma$ Gaussian distributed error expected on the parameter when decomposing a single spectrum. Thus a peak SNR of about 15 corresponds to a deviation or error of 0.1 on the fitted parameter when using this decomposition method. Over most observations in this sample, the \cii\ data have the lowest peak SNR, and therefore their SNR is the main limit on the ability to decompose the \cii\ spectra accurately.

In addition to the accuracy, we are also interested in determining whether a two-component model, using both the CO and \hi\ spectra as templates, gives a statistically better result than a one-component model using either of the templates. This can be thought of as a \textit{nested model}, as the one-component model is a subset of the two-component model (i.e., it is the two-component model with one parameter equal to zero). Adding more parameters to a nested model will always produce a lower $\chi^2$, but the improvement may not be significant.

We compare the possible models through the \textit{F-test} \citep[][ \S~4.6]{mendenhall}. While the F-test is often used in analysis of variance (ANOVA), it can also be used in regression analysis to test whether the simpler of two models provides a better fit. We calculate the F-statistic through:

\begin{equation}
\label{eq:f-test}
    F = \frac{(\chi^2_{\rm{1comp}} - \chi^2_{\rm{2comp}})/(q-p)}{\chi^2_{\rm{2comp}}/(N-q)}
\end{equation}
where $\chi^2_{\rm{1comp}}$ is the $\chi^2$ for the simpler model, $p$ is the number of parameters in the simpler model,  $\chi^2_{\rm{2comp}}$ is the $\chi^2$ for the complex model, $q$ is the number of parameters in the complex model, and $N$ is the number of data points.
The F statistic defined in \autoref{eq:f-test} follows the F-distribution with $(q - p, N - q)=(1,N-2)$ degrees of freedom. We define a null hypothesis that the more complex model does not provide a significantly better fit than the simpler model. We can reject this null hypothesis, implying that the complex model provides a better fit, when the F-statistic is greater than a given critical value from the corresponding F-distribution. 

We show in \autoref{fig:unique} the median F-statistic percentile for the same range of $w_{\rm{CO}}$ and peak SNR as used in the accuracy simulations. The one-component model is defined by fitting the CO and \hi\ template to the simulated data and selecting the fit with the lowest $\chi^2$ value. The F-statistic percentile is dependent on the value for $w_{\rm{CO}}$, since it is easier to see the effect of both components when they contribute approximately equally (note that $w_{\rm{HI}} = 1 - w_{\rm{CO}}$). We cannot statistically distinguish between the two-component model and a one-component model with a peak SNR of five. A peak SNR of ten does a better job, but only for $w_{\rm{CO}} = 0.4-0.6$. The spectra therefore need to have a high SNR of at least fifteen to distinguish between a one-component and a two-component model for cases where the lesser component contributes 20\% or more of the signal. The majority of the individual \cii\ spectra from M101 and NGC~6946 have a peak SNR of less than ten. Combined with the simulations on the accuracy of the decomposition method, we conclude we need a higher peak \cii\ SNR than that provided by most individual spectra in this sample: we achieve this through averaging the data (see \S\ref{sec:stack}).

\section{Results}\label{sec:decomp_all}
In order to identify the dominant phase of the ISM traced by \cii , we compare the velocity resolved profiles of \cii\ from the SOFIA/GREAT data to the profiles of \hi\ 21~cm emission, a tracer of the atomic phase, and to CO~$\rm J=2-1$ emission, a tracer of the molecular phase. The physical spatial resolution of the \cii , CO, and \hi\ data is $\sim$500~pc for M101 and NGC 6946. The profiles carry information about the bulk motions of the given gas phase at this resolution, and we will use their shape to identify the origin of the \cii\ emission. Early detections of \cii\ emission from line-of-sight observations of the Milky Way revealed the multi-phase and extended nature of \cii\ emission \citep{Stacey1985,Shibai1991, Wright1991,Madden1993,Bennett1994, Makiuti2002}. In order to quantify the amount of \cii\ that is associated with each phase in the plane of the Milky Way, velocity resolved spectra are required, as performed by \cite{pineda2013}. We apply a similar velocity resolved approach to decompose the \cii\ emission into the component phases in two nearby galaxies outside of the local group. The multi-phase nature of the \cii\ emission can be inferred by inspection of \autoref{fig:ex_spec} (especially region N5-6), where the \cii\ profile widths are intermediate between those for CO and HI.

\subsection{Linewidth comparison}
\begin{figure}
\centering
\includegraphics[width=\columnwidth]{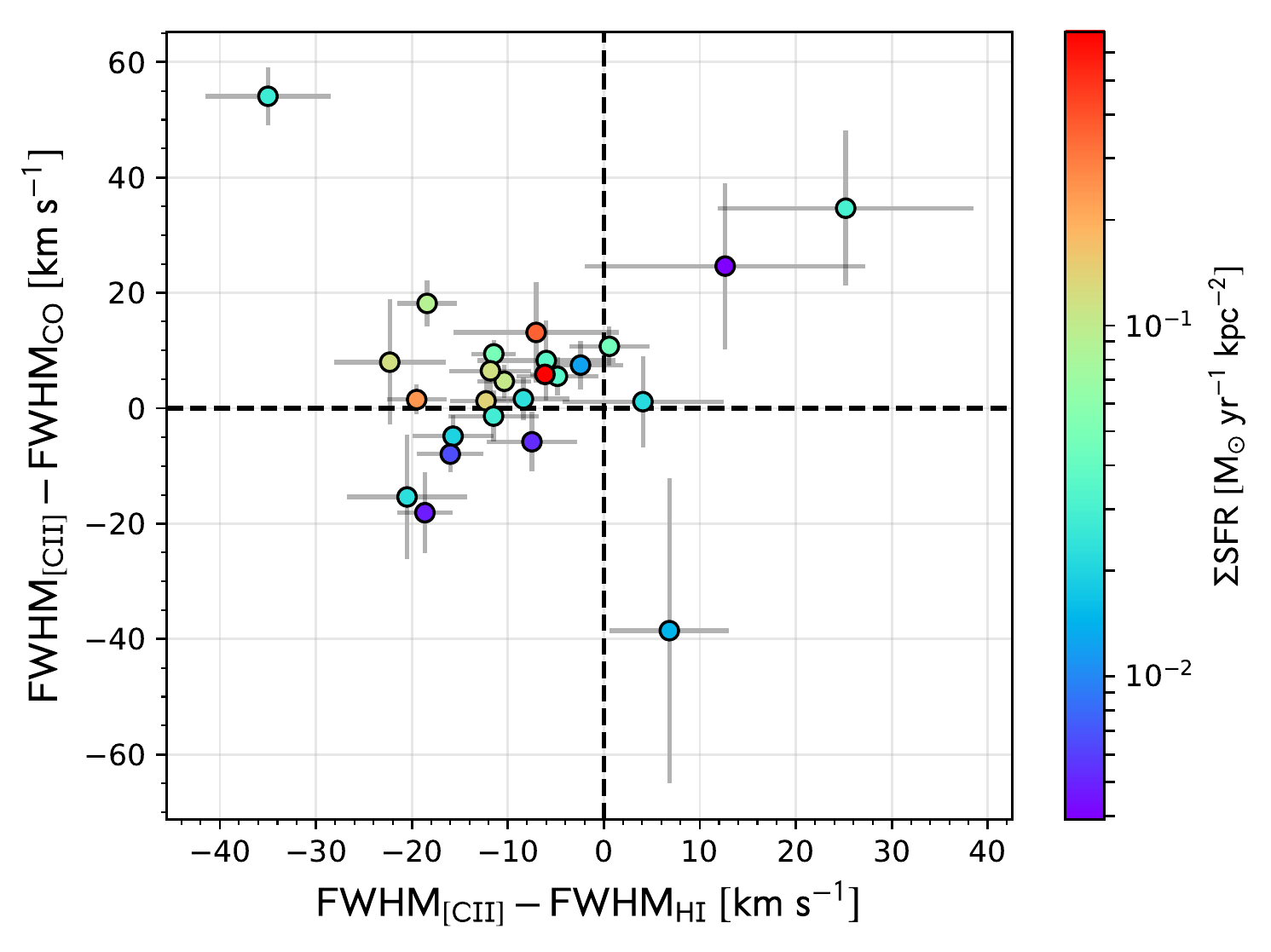}
\caption{The linewidth of the fitted Gaussian curves for the \hi, CO, and \cii\ data as traced by the FWHM. Points to the left of the zero line mean that the HI line profiles are wider than the \cii\ while points above the zero line show \cii\ line profiles that are wider than the CO. Most of the \cii\ profiles have a width in between the CO and HI (see upper left quadrant), suggesting a combined origin of the two.}
\label{fig:gau}
\end{figure}

By fitting the spectral line profiles of the \cii, CO, and \hi\ data, we can quantify how the \cii\ emission is intermediate between the CO and \hi\ emission. We fit each spectra with a Gaussian profile and compare the linewidths between the spectra by examining the fitted full width at half maximum (FWHM) of each line. \autoref{fig:gau} shows the difference between the FWHM of the \cii\ and the CO profile (on the y-axis) or the \hi\ profile (on the x-axis). Most points are found in the fourth quadrant, indicating that the \cii\ FWHM lies between those of CO and \hi. There are no strong trends with \sSFR , shown through the colorscale, and the difference between the FWHM of the \cii , CO, and \hi . The mean FWHM for the \cii\ is 27.7\,\kms\ while the mean FWHM for CO and \hi\ are 22.1\,\kms\ and 35.7\,\kms, respectively. On average, the \hi\ FWHM is 29\% wider than the \cii\ profile and the CO FWHM is 25\% narrower than the \cii\ profile. Other studies have also shown linewidth differences between the \cii, CO, and \hi\ spectra, with up to a 50\% difference between the CO and \cii\ profiles \citep[e.g.][]{deBlok2016, requena2016, Lebouteiller2019}. 

\changes{Gaussian curves fit most of these spectra well, but there are some instances where the fit is poor ($\rm \chi^2_{red} \simeq 2.5$), often when spectra are not symmetric or have a lower peak SNR. These asymmetries provide motivation for using a \cii\ decomposition method that does not assume a line shape (see \S\ref{sec:decomp}). Additionally, the poor Gaussian fits are the points in \autoref{fig:gau} that have higher errorbars.}

Note that the kinematic decomposition method would not be effective if the CO and \hi\ profiles are too similar. The \hi\ spectra, however, are on average 62\% wider than the CO spectra. Thus the tracers of the molecular and atomic gas are sufficiently different to provide an accurate decomposition of the \cii\ emission (see \S\ref{sec:eval}). Further, the difference between the spectral profiles of \cii , CO, and \hi\ are also larger than the velocity resolution of these data. 

\startlongtable 
\begin{deluxetable}{cccccc} 
\tabletypesize 
\footnotesize 
\tablecaption{\cii\ Gaussian Fits Summary \label{tab:gau_tab}} 
\tablehead{ 
\colhead{Galaxy} & \colhead{Region} &  \colhead{A}  & \colhead{v$_\mathrm{peak}$} & \colhead{FWHM}
\\ & & \colhead{(K)} & \colhead{(\kms)} & \colhead{(\kms)}}
\startdata 
M101&Ma&0.95$\pm$0.04&273.6$\pm$0.6&27.4$\pm$1.3 \\
M101&M3-0&0.20$\pm$0.03&201.6$\pm$2.0&31.1$\pm$4.8 \\
NGC6946&Nd&0.61$\pm$0.06&110.5$\pm$1.7&33.1$\pm$4.0 \\
NGC6946&N5-5&0.24$\pm$0.02&-21.7$\pm$1.4&30.1$\pm$3.3 \\
\enddata 
\tablecomments{The Gaussian fitted parameters of the [CII] lines for the 3$\sigma$ sample. A represents the amplitude of the Gaussian, $\rm{v_{peak}}$ is the fitted peak velocity, and FWHM is the full width half maximum of the Gaussian fit. (This table in its entirety is available in a machine-readable form online.)} 
\end{deluxetable}

\subsection{Contributions from ionized gas}
\label{sec:ionizedgas}
The \nii\ 205\,\micron\ transition arises from ionized gas because nitrogen has an ionization potential of 14.5\,eV, greater than that of hydrogen, and can be used to isolate the contribution the ionized gas has on the \cii\ emission. The similar critical densities for collisions with electrons, $n_e \approx32$\,\density\ for \nii\ 205\,\micron\ and $n_e \approx45$\,\density\ for \cii\ \citep[]{schoier2005}, mean that the \cii/\nii\ line ratio has a weak dependence on the density and ionization state. Therefore, for a given $\rm N^+/\rm C^+$ abundance ratio, the observed \cii/\nii\ line ratio gives a relatively density-independent estimate of the contribution of ionized gas on the total \cii\ emission \citep{Oberst2006}. 

\nii\ 205\,\micron\ is a faint line compared to \cii, and consequently all the \nii\ observations we have from SOFIA/GREAT are non-detections. We use the 3$\sigma$ rms of \nii\ to compute a lower limit on the \cii/\nii\ ratio. This ratio can then be used to find a lower limit on \fneut, the fraction of molecular and atomic gas that contributes to the overall \cii\ intensity (or, conversely, an upper limit to the fraction of emission contributed by the ionized gas). We compare the observed \cii/\nii\ ratio to the theoretical ratio derived from the ionic abundance of $\mathrm{C^+/N^+}$ and attribute any excess to the contribution the neutral gas has to the \cii\ emission. The theoretical \cii158\,\micron/\nii205\,\micron\ ratio does depend slightly on density, but ranges between 3.1 at low densities and 4.2 for high densities \citep{Oberst2006}. We use a \cii158\,\micron/\nii205\,\micron\ ratio of 4, the same as \citet[]{Croxall2017}, in order to compare to their results. This value comes from calculations of the collision rates of $\rm e^-$ with $\rm C^+$ \citep[]{Tayal2008} and $\rm N^+$ \citep{Tayal2011} and assumes Galactic gas phase abundances for carbon ($\mathrm{X_{C/H} = 1.6 \times \ 10^{-4}}$, \citealt[]{Sofia2004}) and nitrogen ($\mathrm{X_{N/H} = 7.5 \times \ 10^{-5}}$, \citealt[]{Meyer1997}). We calculate \fneut\ by subtracting the ionized gas contribution to \cii:

\begin{equation}
f_{\rm{neutral}} = \frac{I_{\rm{[CII]}} - R_{\rm{ionized}} \times (3 \mathrm{\sigma_{rms,[NII]}})}{I_{\rm{[CII]}}}
\end{equation}
where $R_{\rm{ionized}}$ = 4, the approximate theoretical \cii/\nii\ ratio. 

\begin{figure}
\centering
\includegraphics[width=\columnwidth]{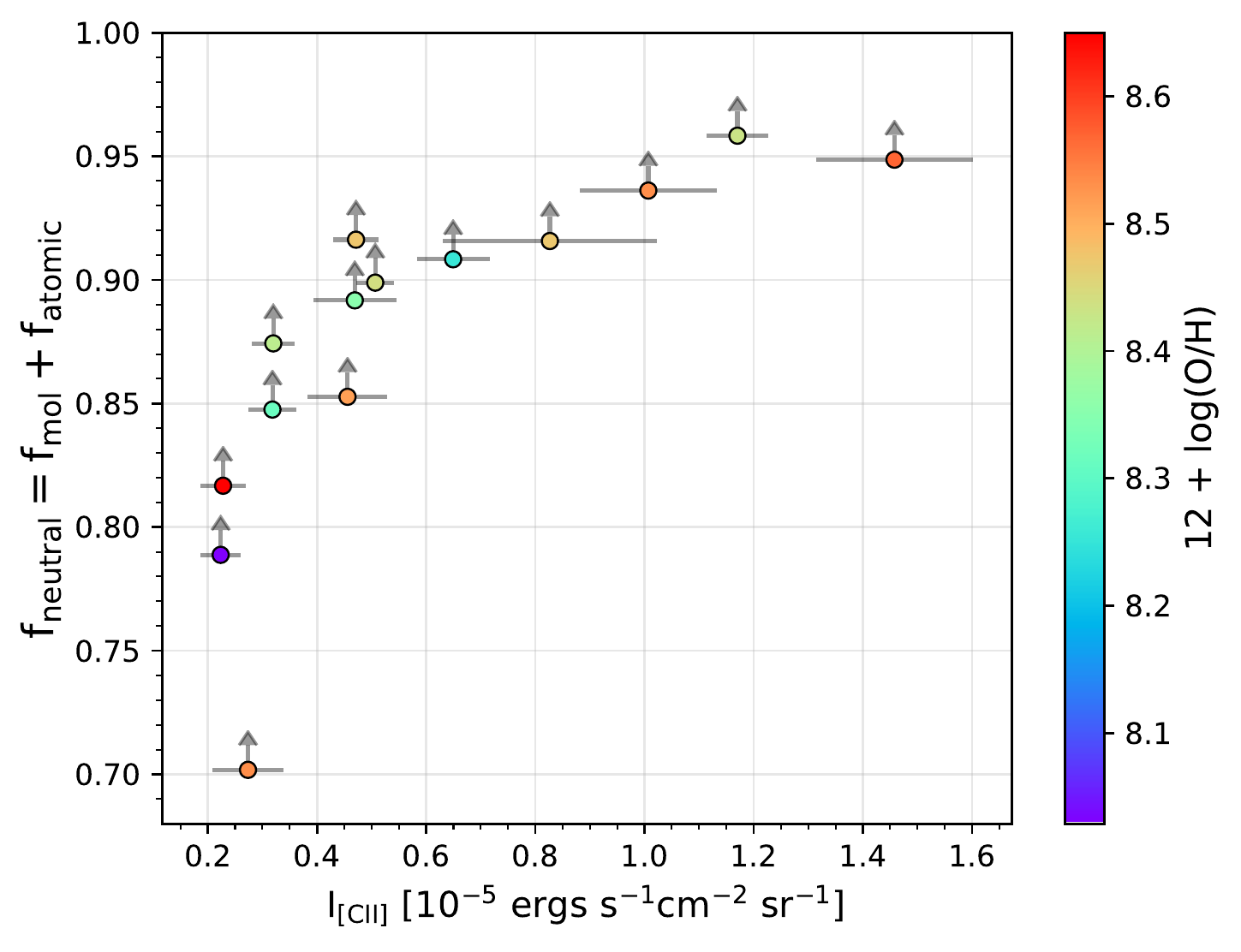}
\caption{Lower limits of the contribution the neutral gas has to the \cii\ emission found through 3$\sigma$ rms \nii\ 205~\micron\ measurements. Colors represent the gas phase oxygen abundances. The average limit \fneut\ $\gtrsim$ 88\%, suggesting the the contribution of ionized gas to the \cii\ emission is negligible.}
\label{fig:NII}
\end{figure}

We present the \fneut\ lower limits in \autoref{fig:NII}, with the colorscale representing the oxygen abundance we estimate from the metallicity gradients found in \citet[]{Pilyugin2014}. All of the limits show an \fneut\ greater than 70\% with an average of \fneut\ = 88\%. There is also a trend with the \cii\ intensity, suggesting that the regions with brighter \cii\ have a smaller possible contribution from the ionized gas. 

The estimation of \fneut\ assumes a $\mathrm{C^+/N^+}$ ratio, which we anchor to the Galactic $\mathrm{C/N}=2.13$ ratio at log(O/H) $\approx$ 8.65 \citep{Simon-Diaz2011}. There is an expected variation of the $\mathrm{C/N}$ ratio with metallicity \citep{Nieva2012}, but \citet[]{Croxall2017} show that \fneut\ varies by only 10\% for a change of 0.8 dex in oxygen abundance. Therefore our original calculation using a Galactic abundance will only marginally change the already minimal contribution the ionized gas has to the \cii\ emission. 

The \fneut\ lower limits we compute are very similar to the work by \cite{Croxall2017}, who find (74$\pm$8)\% of the \cii\ emission comes from the neutral gas in galaxies from KINGFISH \citep[]{Kennicutt2011}. A similar result is found for regions in the LMC, where \fneut~$\gtrsim$~90\% \citep[]{Lebouteiller2019}, and in the measurements of low metallicity galaxies in the Dwarf Galaxy Survey, which estimate \fneut\, $>$\,70\% \citep{cormier2019}. According to these limits, we assume that the contribution of the ionized gas to the \cii\ emission is negligible. This allows us to spectrally decompose the \cii\ emission using only tracers for molecular and atomic gas. 

\begin{figure*}
\centering
\includegraphics[width=1\textwidth]{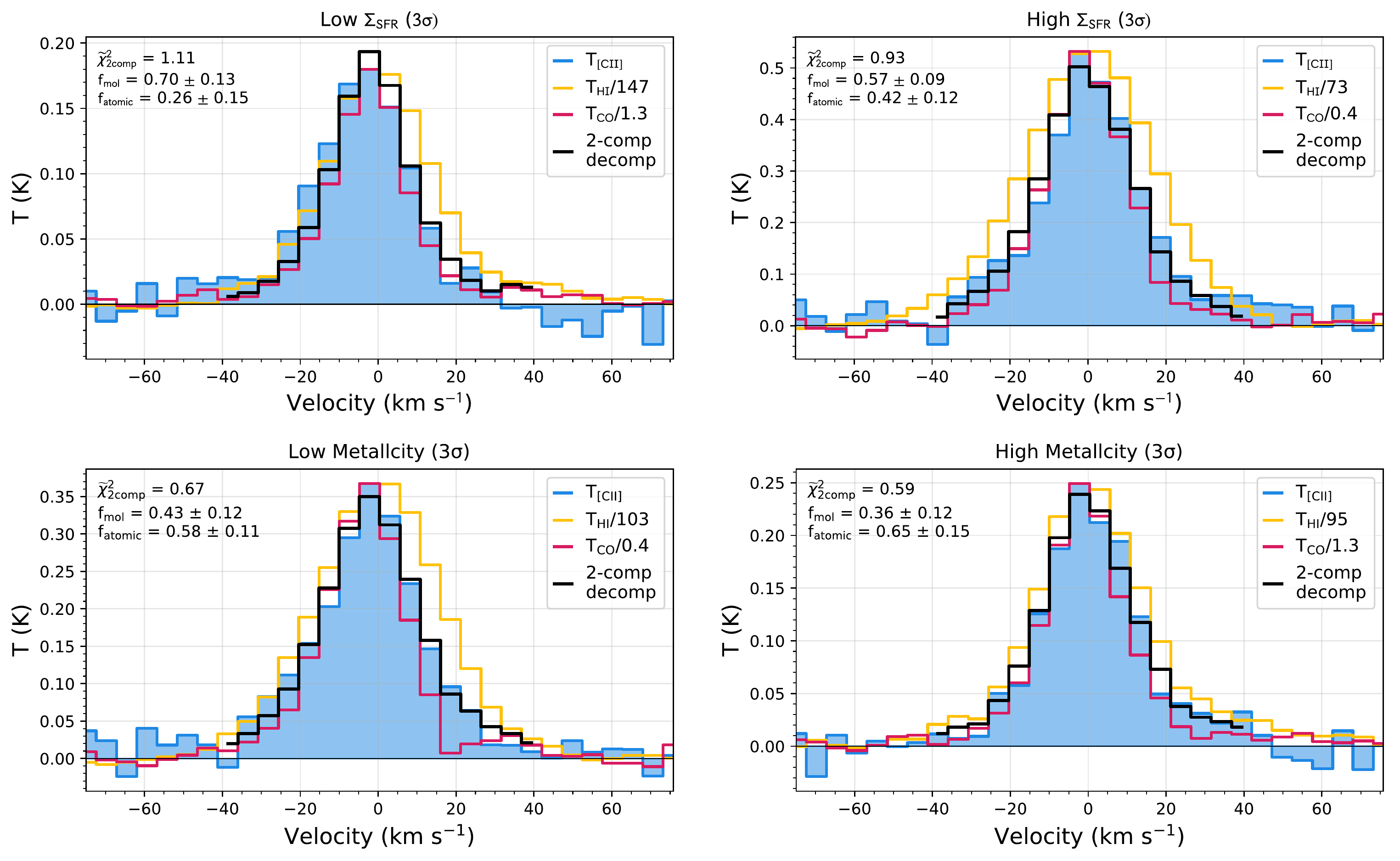}
\caption{Decomposition of stacked spectra from the 3$\sigma$ sub-sample (includes all regions with integrated \cii\ intensities three times greater than the rms noise). The top two panels show the sample split by \sSFR\ where the division is \SFRcut . The bottom two panels show the sample split by metallicity, where the division is 12 + log(O/H) = 8.55. Blue colors represent the stacked \cii\ spectrum, yellow corresponds to the stacked \hi\ spectrum, and red shows the stacked CO data. The CO and \hi\ spectra are scaled to the amplitude of the \cii\ spectrum. The black line is the two component decomposition using the scaled HI and CO data to best reproduce the \cii\ spectrum. The reduced $\chi^2$ of the two component fit and the results of the decomposition are shown in the upper left corner. The agreement between the decomposition fit (black) and the original \cii\ spectrum (blue) are generally very good.}
\label{fig:decomp}
\end{figure*}

\subsection{Decomposition of spectrally averaged data}\label{sec:stack}

As discussed in \S\ref{sec:eval}, we need a high peak signal-to-noise ratio (SNR) of 10-15 to accurately spectrally decompose the \cii\ emission. Only three spectra in this dataset have a peak SNR greater than 10, and these individual spectra are analyzed in  \S\ref{sec:ind}. In order to achieve the SNR required to accurately decompose the \cii\ spectra, we average the bulk of the data by combining similar spectra together through a process called stacking. 

We stack the spectra by aligning them in velocity, with the \hi\ data providing the velocity reference because the \hi\ spectra have a higher SNR compared to the CO spectra. First, we shift all of the spectra to the center of the \hi\ Gaussian-fitted peak velocity. We then interpolate over a velocity grid with a 150\,\kms\ bandwidth and the 5.2\,\kms\ velocity resolution. The centered, interpolated spectra for the \cii, CO, and \hi\ are averaged together. This method reduces the noise of the final spectrum by a factor of $\sim\sqrt{N_{spec}}$, where $N_{spec}$ is the number of spectra stacked in the given cut and sample. The peak SNR of the stacked \cii\ spectra thus increases to about 10 or higher, making a meaningful spectral decomposition possible. For example, stacking 13 spectra together each with an average SNR\,$\sim$\,5 produces a stacked spectrum with an SNR\,$\sim$\,15. 
 
In the process of stacking the data, we wish to preserve any relation between the environmental properties in a region and the results of the \cii\ decomposition. For this study, we will stack spectra binning by star formation rate surface density (\sSFR), metallicity (Z), \changes{and normalized galactocentric radius (\gc)}. We bisect the data into a ``low" \sSFR\ bin and a ``high" \sSFR\ bin, with a cutoff value of \SFRcut . We use the same process for metallicity by defining a cutoff value of 12 + log(O/H) = 8.55 \changes{and for the galactocentric radius with a cutoff value of \gc\ = 0.4}. These values were chosen by splitting the 3$\sigma$ sample of \cii\ spectra into roughly equal numbered bins. The minimum metallicity for the 3$\sigma$ sample is 12 + log(O/H) = 8.03 and the maximum is 12 + log(O/H) = 8.65. The \sSFR\ varies between $3.9 \times 10^{-3}$ and $6.9 \times 10^{-1}$ \msun\ yr$^{-1}$ kpc$^{-2}$. \changes{These spectra also span the distribution of galactocentric radii, with \gc\ ranging from 0.1 to 0.9.} In addition, we stack all the spectra from M101 and NGC~6946 separately in order to identify whether there are differences in the results of the \cii\ decomposition in a given galaxy. Lastly, we produce a stack of all spectra together regardless of the property or galaxy (labeled ``all spec"). 

Because the results of the decomposition may depend on the sub-sample of \cii\ spectra used, we produce different stacks with selections as described in \S\ref{sec:sample}. We include spectra by using three different criteria: a 3$\sigma$ sub-sample (defined as including spectra where the \cii\ integrated intensity of a given \cii\ spectrum is 3 times the rms noise), an intensity sub-sample (defined as including spectra when the \cii\ integrated intensity is greater than 5\,K\,\kms), and the sample where all the \cii\ spectra are selected, including non-detections (labeled ``no cuts''). The 3$\sigma$ and I$>$5\,K\,\kms\ intensity sub-samples give similar results; therefore we present stacked data using just the 3$\sigma$ and ``no cut'' samples. 

We then fit the stacked spectra with a two-component model, defined as the linear combination of the CO and HI data, as well as a one-component model, which uses only the CO or HI data as templates. We compare the goodness of fit of the one-component model with the lowest $\chi^2$ to the two-component model using the F-test, as described in  \S\ref{sec:eval}. We find that a two-component model fits the data statistically better in most instances, except for stacks corresponding to the low \sSFR\ bin, which has a lower \cii\ SNR than other stacks, and the stacks where both models provide a bad fit ($\rm \chi^2_{red} \approx 3.2$). 

\startlongtable 
\begin{deluxetable*}{cccccccccccccc} 
\tabletypesize 
\footnotesize 
\tablecaption{Results of the decomposition\label{tab:decomp_tab}} 
\tablehead{ 
\colhead{Property} & \colhead{Sample} & \colhead{N$_{\rm spec}$} & \colhead{SNR$_{\rm [CII]}$} & \colhead{$\mathrm{\Sigma}$SFR} & \colhead{12 + log(O/H)} &   \colhead{R/$\mathrm{R_{25}}$} & \colhead{$\mathrm{\Sigma_{gas}}$} & \colhead{f$_{\rm mol}$} & \colhead{f$_{\rm atomic}$} & \colhead{$\widetilde{\chi}^2_{2c}$} & \colhead{$\widetilde{\chi}^2_{1c}$}  } 
\startdata 
High $\mathrm{\Sigma}$SFR&3$\sigma$& 10&16.62&20.01&8.42&0.52&58.31&0.57$\pm$0.09&0.42$\pm$0.12&0.93&1.57 \\
High $\mathrm{\Sigma}$SFR&No Cuts& 10&16.62&20.01&8.42&0.52&58.31&0.57$\pm$0.09&0.42$\pm$0.12&0.93&1.57 \\
Low $\mathrm{\Sigma}$SFR&3$\sigma$& 17&12.15&1.69&8.59&0.38&45.11&0.70$\pm$0.13&0.26$\pm$0.15&1.11&1.15 \\
Low $\mathrm{\Sigma}$SFR&No Cuts& 68&10.89&0.77&8.50&0.46&27.27&0.43$\pm$0.15&0.57$\pm$0.17&1.63&1.80 \\
High Z&3$\sigma$& 13&15.71&1.66&8.62&0.25&48.51&0.36$\pm$0.12&0.65$\pm$0.15&0.59&0.80 \\
High Z&No Cuts& 27&15.73&1.01&8.62&0.23&36.84&0.36$\pm$0.15&0.65$\pm$0.19&0.47&0.67 \\
Low Z&3$\sigma$& 14&16.58&13.63&8.42&0.58&43.88&0.43$\pm$0.12&0.58$\pm$0.11&0.67&2.34 \\
Low Z&No Cuts& 51&18.43&4.10&8.40&0.59&26.24&0.21$\pm$0.10&0.81$\pm$0.14&3.29&3.27 \\
High R/$\mathrm{R_{25}}$&3$\sigma$& 13&14.86&10.91&8.42&0.60&49.60&0.52$\pm$0.09&0.48$\pm$0.11&0.68&1.34 \\
High R/$\mathrm{R_{25}}$&No Cuts& 45&14.84&3.49&8.40&0.64&28.03&0.47$\pm$0.14&0.53$\pm$0.13&0.76&1.21 \\
Low R/$\mathrm{R_{25}}$&3$\sigma$& 14&12.89&6.38&8.61&0.26&49.84&0.47$\pm$0.14&0.53$\pm$0.16&0.60&1.01 \\
Low R/$\mathrm{R_{25}}$&No Cuts& 33&10.13&2.97&8.58&0.25&35.97&0.47$\pm$0.23&0.53$\pm$0.26&1.39&1.44 \\
M101&3$\sigma$& 13&21.30&12.98&8.48&0.37&38.32&0.28$\pm$0.09&0.69$\pm$0.18&1.28&2.74 \\
M101&No Cuts& 40&11.40&3.73&8.45&0.37&23.65&-0.03$\pm$0.09&1.03$\pm$0.14&3.20&2.99 \\
NGC 6946&3$\sigma$& 21&15.66&5.96&8.56&0.51&56.44&0.59$\pm$0.10&0.41$\pm$0.11&0.32&0.74 \\
NGC 6946&No Cuts& 45&16.31&2.86&8.52&0.56&38.37&0.63$\pm$0.15&0.38$\pm$0.12&0.35&0.64 \\
All Spec &3$\sigma$& 27&23.04&8.56&8.53&0.46&49.73&0.48$\pm$0.07&0.52$\pm$0.07&0.40&1.97 \\
All Spec &No Cuts& 78&23.73&3.27&8.49&0.48&31.39&0.34$\pm$0.09&0.66$\pm$0.11&1.11&1.83 \\
\enddata
\tablecomments{The results of the stacked decomposition. The \sSFR\ bin corresponds to the star formation rate surface density with a low/high cutoff value of \SFRcut , the Z bin corresponds to the 12 + log(O/H) with a low/high cutoff value of 12 + log(O/H) = 8.55, \changes{and the R/$\mathrm{R_{25}}$ bin is the normalized galactocentric radius with a low/high cutoff value of R/$\mathrm{R_{25}} =0.4 $}. Sub-Sample represents how the regions were selected, with 3$\sigma$ representing regions with integrated [CII] intensities three times greater than the RMS noise and No Cuts containing all of the spectra in the data. The average weighted \sSFR\ (in units of $10^{-2}$ \msun\ yr$^{-1}$ kpc$^{2}$), metallicity in 12 + log(O/H), \changes{normalized galactocentric radius ($\mathrm{R/R_{25}}$), and the total gas surface density ($\rm \Sigma_{gas} = \Sigma_{H{\small I}} + \Sigma_{H_2}$) in units of $\rm \msun \  pc^{-2}$ of the stacked spectra are reported}. $f_{\rm atomic}$ and $f_{\rm mol}$ are the fraction of the integrated [CII] intensity that is associated with the atomic and molecular gas, respectively. $\widetilde{\chi}^2_{\rm 2comp}$ is the reduced $\chi^2$ for the model using both HI and CO while $\widetilde{\chi}^2_{\rm 1comp}$ is the reduced $\chi^2$ for the CO or HI one-component model that has the smallest $\chi^2$}
\end{deluxetable*}

\begin{figure*}
\centering
\includegraphics[width=1\textwidth]{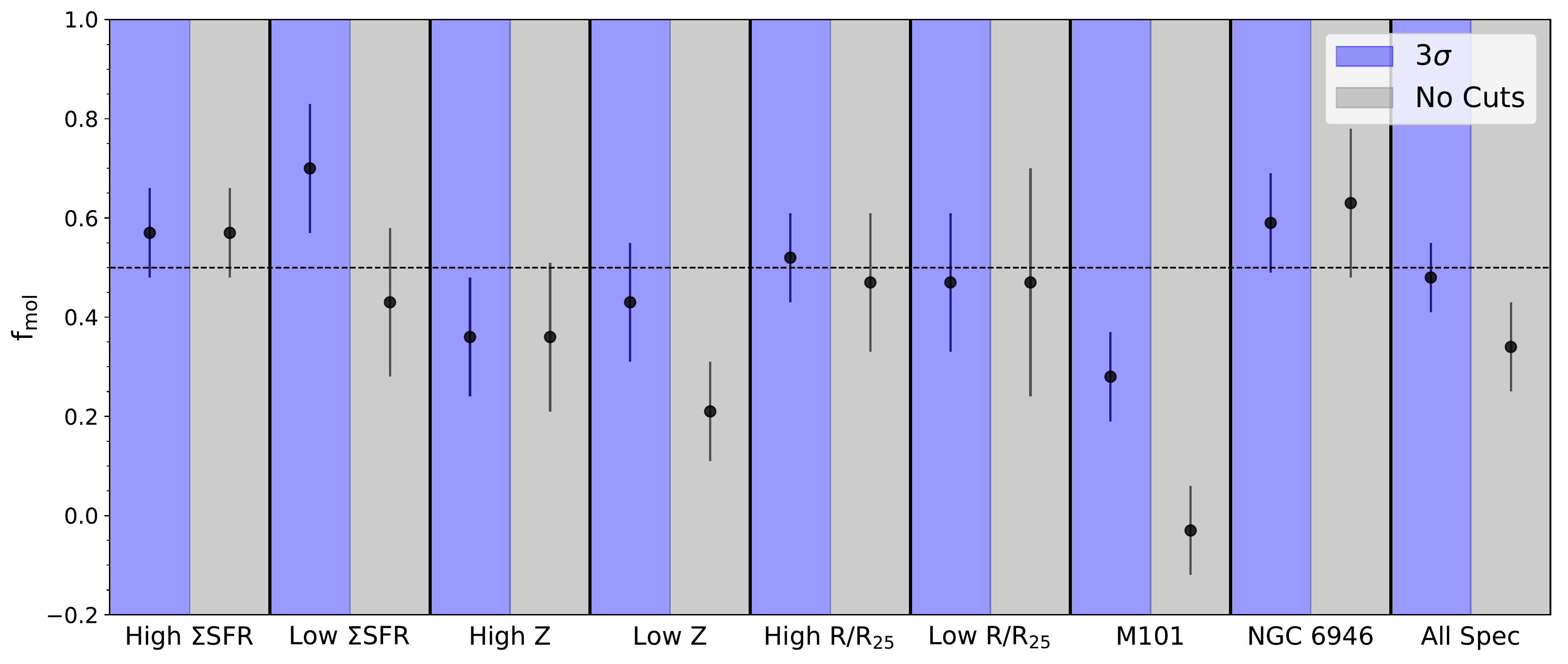}
\caption{Summary of the \cii\ decomposition of averaged spectra. The stacked property is named on the x-axis and the sub-sample used is colored in blue for the 3$\sigma$ sample and gray for when no cuts are made. Most stacked bins contain \cii\ with an equal or greater contribution from the atomic gas. There is also a slight trend of decreasing \fmol\ when the ``no cuts'' sub-sample is used. The spectra from NGC~6946 are more dominated by the molecular gas when compared to the spectra in M101.}
\label{fig:decomp_sum}
\end{figure*}

\autoref{fig:decomp} shows an example of the stacked \cii , \hi , and CO spectra for the \sSFR\ and metallicity property bins as well as the resulting \cii\ spectral decomposition. The full results are given in \autoref{tab:decomp_tab} and a summary of the decomposition results for \fmol\ is plotted in \autoref{fig:decomp_sum}. We calculate the average \sSFR,  metallicity, \gc , and \changes{total gas surface density} for each bin and report the result in \autoref{tab:decomp_tab}. \changes{We use the \hi\ and CO line intensities at the $\sim$500~pc resolution and Equations 1 and 2 from \citet[]{herrera2017} to calculate the total gas surface density for these regions.} 

Most of the stacked spectra show that the atomic gas has an equal or larger contribution to the overall \cii\ intensity. For example, the high metallicity (Z) bin yields values of $\fmol = 0.36 \pm 0.12$ and $\fatomic = 0.65 \pm 0.15$, suggesting that the high metallicity points have \cii\ emission that is present slightly more in the atomic phase. Similarly, the high \sSFR\ bin contains \cii\ emission that is equally distributed between the two phases, with a $\fmol = 0.57 \pm 0.09$ and an $\fatomic = 0.42 \pm 0.12$. When taking into account the uncertainties, all stacked bins are approximately consistent with a 50\% or more contribution from the atomic phase to the \cii\ decomposition.

We also find that the fraction of \cii\ coming from the molecular phase decreases or remains the same when comparing the ``no cuts'' sample to the 3$\sigma$ sub-sample. The ``no cuts'' sample uses all of the spectra in the dataset and consequently contains fainter and non-detected \cii\ spectra. When there are more non-detections included in a stacked bin, the atomic gas tends to have a larger contribution to the \cii\ emission. This trend is relatively consistent for the different properties studied, except in cases where \fatomic\ stays constant. 

% Further, the high \sSFR\ bin contains the same spectra in both sub-samples due to the SFR-\cii\ correlation \citep[]{Stacey1991,Boselli2002,DeLooze2014, Herrera2015}. Therefore, the ``no cuts'' sample is not shown in \autoref{tab:decomp_tab} for the high \sSFR\ bin because it is equivalent to the 3$\sigma$ sub-sample. 

There is a clear trend when comparing the stacked bins between the two galaxies. The molecular gas dominates the \cii\ emission significantly more in NGC~6946 than M101. Additionally, the contribution of molecular gas to the \cii\ emission stays consistent for the ``no cut'' sample in NGC~6946 but decreases for M101.

\begin{figure}
\centering
\includegraphics[width=\columnwidth]{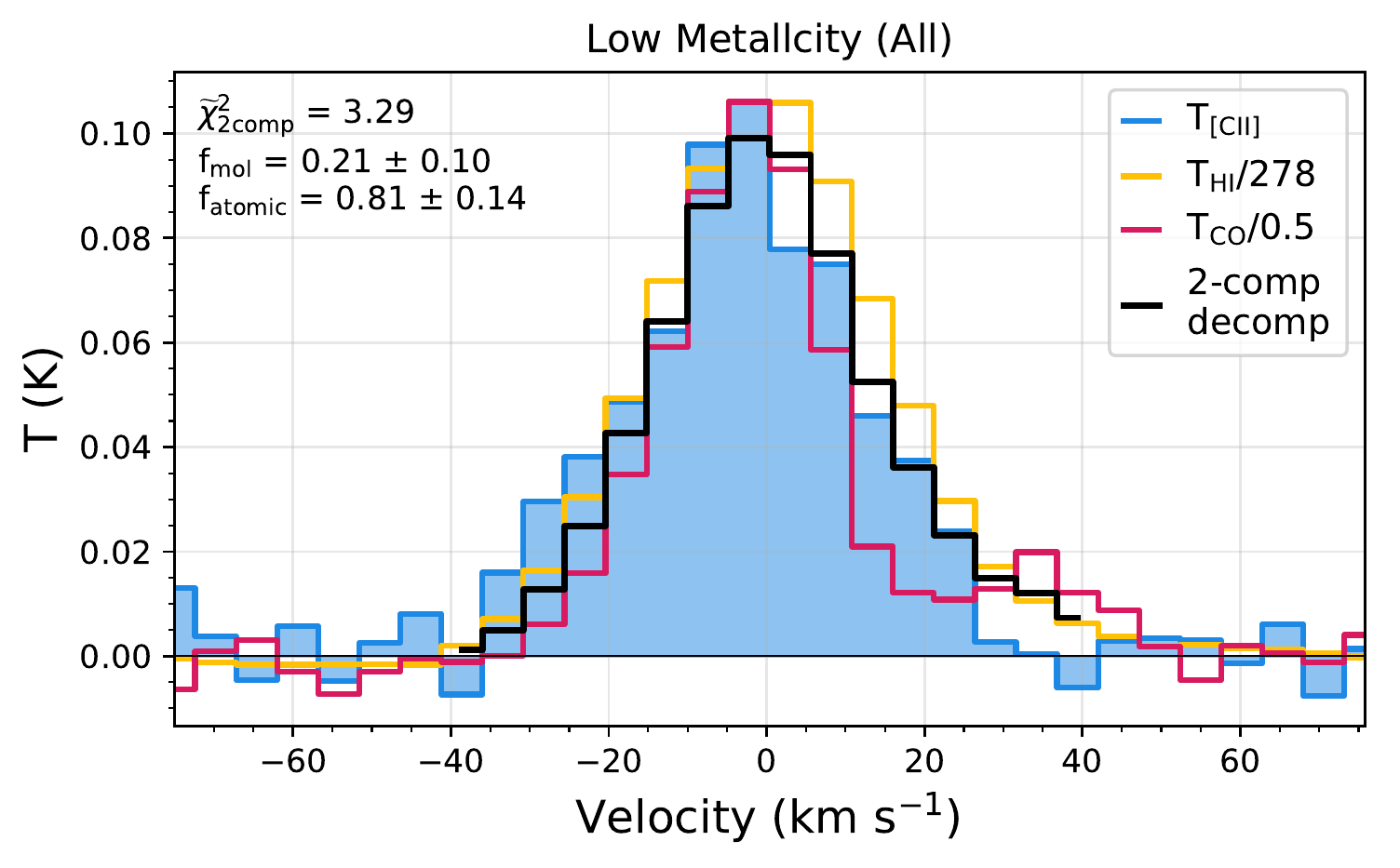}
\caption{Spectra of the low metallicity stacked bin when making no cuts on the included regions. See \autoref{fig:decomp} for description of line colors and labels. There is \cii\ emission on the left wing that isn't traced by the CO or HI data, leading to a high $\chi^2$ for the two-component model. The excess emission may be from CO-dark gas or another contribution not traced in this work. }
\label{fig:lowZ}
\end{figure}

We find high $\chi^2$ values of $\chi^2 = 3.29$ for the low metallicity bin and $\chi^2 = 3.20$ for the M101 bin in the ``no cuts'' sample, suggesting that the two component model does not fit these data well. \autoref{fig:lowZ} shows the stacked spectrum for the low metallicity bin and there is excess \cii\ emission not traced by the two-component model. This suggests that the underlying assumption, that the combination of an atomic phase (as traced by the HI data) and molecular phase (as traced by the CO data) can completely explain the \cii\ emission profile, may not apply. An additional component, either ionized gas (which we think unlikely in view of the discussion in \S\ref{sec:ionizedgas}) or much more likely CO-faint molecular gas, may be needed to appropriately model the \cii\ emission.

At low metallicities, CO-faint molecular gas can be a large contributor to the  \cii\ emission, hinting that the high $\chi^2$ and additional \cii\ emission in \autoref{fig:lowZ} is likely CO-faint gas \citep[e.g.][]{Madden1997, grenier2005, Wolfire2010, Jameson2018}. \citet{fahrion2017} also observe wider wings in their \cii\ profiles, but their origin was not attributed to the CO or \hi\ gas. Similar work using higher resolution data from the Small and Large Magellanic Clouds show wide \cii\ profiles not associated with the atomic, molecular, or ionized gas \citep{okada2015, okada2019, requena2016, Lebouteiller2019}. With the present data, however, we cannot ascertain the exact nature \changes{of} this component. 

There are no clear differences in the \cii\ decomposition when comparing the high and low cuts of the \sSFR\ and metallicity. The spectra come from $\sim$500~pc regions, which may contain multiple PDR and \hii\ complexes, and averaging over the 15\arcsec\ beam can dilute trends with metallicity or star formation rate. The necessity to stack data may further weaken possible trends by averaging over multiple spectra from two different galaxies. Interestingly, the largest difference between the results of the \cii\ decomposition comes from comparing the two galaxies. M101 contains \cii\ emission dominated more by the atomic gas while the \cii\ in NGC~6946 comes more from the molecular gas. The physical property driving the difference between these \cii\ decomposition results is uncertain, but is likely not the aggregate star formation rate or metallicity of these galaxies.

\changes{We also investigate the dependence on galactocentric radius and the \cii\ decomposition and find a similar value of  $\rm f_{atomic} \approx 0.5$ for both the low and high \gc\ cuts. There is no difference in the distribution of galactocentric radius between NGC~6946 and M101, suggesting that \gc\ is not the reason for the varying \cii\ decomposition results in these two galaxies. The galactocentric radius does slightly increase with the ``no cuts'' sample, but because there is no difference between the low and high \gc\ bins, the small increase in \gc\ is unlikely causing the larger contribution of atomic gas to the \cii\ emission seen in the ``no cuts'' sample.}

Lastly, we use bootstrapping techniques to confirm that the process of stacking the spectra is accurate. We randomize the regions included in each bin, then decompose the summed random combination of spectra for that bin, and repeat this process for 500 trials. The results of the bootstrap technique are very similar to the original results, suggesting that the uncertainties in the original decomposition are accurate and that one given spectrum in a bin does not dominate the \cii\ decomposition. Additionally, we do not change the weighting of the original stacked spectra, which are by nature of summing the data fluxed-weighted.  We do not use rms weighting because the brighter \cii\ spectra tend to have shorter integration times and larger rms values.

\subsection{Decomposition of individual spectra}\label{sec:ind}
There are three spectra in our dataset that have a sufficient SNR to produce a meaningful decomposition, as determined by our analysis in \S\ref{sec:eval}. We show the results in \autoref{tab:tab_ind}. Within the uncertainties, the results from these regions agree with those from stacked spectra.

All three data points are in the low metallicity property bin, but \fatomic\ ranges from 0.65 to 0.46, suggesting there may be a fair amount of scatter in the decomposition of individual spectra that make up the bins described in \S\ref{sec:stack}. The individual variation is also seen in similar \cii\ decomposition work by \citet{Mookerjea2016}.

\startlongtable 
\begin{deluxetable*}{cccccccccccc} 
\tabletypesize 
\footnotesize 
\tablecaption{Single Spectrum Decomposition \label{tab:tab_ind}
} 
\tablehead{ 
\colhead{Galaxy} & \colhead{Region} & \colhead{Cycle} & \colhead{SNR$\rm_{[CII]}$} & \colhead{$\mathrm{\Sigma}$SFR} & \colhead{12 + log(O/H)} & \colhead{R/$\mathrm{R_{25}}$} & \colhead{$\mathrm{\Sigma_{gas}}$} & \colhead{f$\rm_{mol}$} & \colhead{f$\rm_{atomic}$} & \colhead{$\rm \widetilde{\chi}^2_{2comp}$} & \colhead{$\rm\widetilde{\chi}^2_{1comp}$} } 
\startdata 
M101&Ma-0&  2&14.84&68.80&8.43&0.33&131.84&0.39$\pm$0.13&0.65$\pm$0.12&1.61&2.25 \\
NGC6946&N2-0&  4&9.20&10.30&8.47&0.73&49.85&0.53$\pm$0.15&0.46$\pm$0.14&0.69&1.12 \\
NGC6946&N3-0&  4&11.16&4.84&8.44&0.83&44.37&0.44$\pm$0.11&0.54$\pm$0.08&1.05&2.51 \\
\enddata 
\tablecomments{The decomposition results for the three individual spectra with the highest SNR$_{\rm [CII]}$. The table headers are the same as \autoref{tab:decomp_tab}.}
\end{deluxetable*}

\section{Discussion}\label{sec:discussion}

\subsection{Thermal Pressure in the Cold Neutral Medium}
\label{sec:pressure}
The spectral \cii\ decomposition allows us to separate \cii\ emission that is directly associated with the atomic gas. Using a method proposed by \citet[]{kulkarni1987} and demonstrated by \citet[]{herrera2017}, we use the \cii\ cooling rate to estimate the thermal pressure in the cold neutral medium (CNM). The thermal pressure is important in determining the cooling curve and pressure equilibrium in the atomic medium \citep{field1969}, has consequences for the amount of cold dense material available for star formation, and it is part of the cycle of self-regulation of star formation activity in galaxies \citep[]{ostriker2010, Kim2011}. 

In order to estimate the thermal pressure in the CNM, we need to relate the observed \cii\ emission to the physical properties of the gas that emits the \cii. The integrated intensity of \cii\ for collisional excitation in the optically thin limit with a given collisional partner is \citep{crawford1985, goldsmith2012} 

\begin{equation}
\begin{split}
    I_{\rm [CII]} =\left( \frac{2e^{-91.2/T}} {1 + 2e^{-91.2/T} + A_{ul}/(\Sigma R_{ul,i} n_i)} \right) N_{\rm C^+} \\ \times \  2.3 \times 10^{-21}\,,
\end{split}
\label{eq:craw}
\end{equation}
where $I_{\rm [CII]}$ is the integrated \cii\ intensity in units of erg\,s$^{-1}$\,cm$^{-2}$\,sr$^{-1}$, $T$ is the kinetic temperature of the collisional partner in K, $N_{\rm C^+}$ is the column density of ionized carbon in the line of sight in units of cm$^{-2}$, $A_{ul}$ is the spontaneous decay rate of the 158\,\micron\ \cii\ transition ($A_{ul}$ = 2.3$\rm \times \ 10^{-6}$\,s$^{-1}$), $n$ is the number density of the collisional partner in units of cm$^{-3}$, and $R_{ul}$ is the collisional de-excitation rate coefficient of a given partner at a kinetic temperature $T$. The sum in the denominator is over all the relevant collisional partners, including $\mathrm{H^0}$, \htwo, He, or $\mathrm{e^-}$. The focus in this section is collisions with the atomic gas, $\mathrm{H^0}$, where $R_{ul}$ is calculated by \citet[]{goldsmith2012}

\begin{equation}
    R_{ul}(H^0) = 4.0 \times 10^{-11}(16 + 0.35T^{0.5} + 48T^{-1})
\end{equation}
and is in units of $\mathrm{cm^3 ~ s^{-1}}$. For $T = 100\,\rm K$, the collisional de-excitation rate coefficient for atomic hydrogen is $R_{ul}(H^0) = 8 \rm \times 10^{-10}\,\mathrm{cm^3 ~ s^{-1}}$. The atomic gas will also contain helium, which has a collisional de-excitation rate coefficient equal to 0.38 times the rate for atomic hydrogen \citep{Draine2011}.

The total observed \cii\ integrated intensity has components from the neutral gas (including the molecular and atomic phases) and the ionized gas. 

\begin{equation}
    I^{\rm tot}_{\rm [CII]} = I^{\rm neutral}_{\rm [CII]} + I^{\rm ionized}_{\rm [CII]},
\end{equation}

\noindent where $I^{\rm neutral}_{\rm [CII]} = I^{\rm atomic}_{\rm [CII]} + I^{\rm mol}_{\rm [CII]}$. We define $f_{\rm{ion}}$ as the fraction of $I^{\rm tot}_{\rm [CII]}$ that comes from the ionized gas.
As part of the neutral phase there is diffuse ``CO-dark'' molecular gas phase that is mixed within the CNM \citep{grenier2005, Wolfire2010, Langer2014}. The kinematics of the CO-dark gas may match those of the CNM and thus the contribution needs to be removed. We define $f_{\rm H_2}$ as the fraction of the \cii\ intensity that originates in the CO-dark molecular gas. 
Therefore we can write

\begin{equation}
I^{\rm atomic}_{\rm [CII]} = (1-f_{\rm H_2})(1-f_{\rm ion})(1-f_{\rm mol})I^{\rm tot}_{\rm [CII]},
\label{eq:I}
\end{equation}

\noindent where we use $f_{\rm mol}$ calculated from the fitting described in \S\ref{sec:stack} to find the fraction of the total \cii\ intensity that originates from the atomic gas alone.

The warm neutral medium (WNM) in the atomic phase has a combination of physical conditions (T $\approx$ 8000 K, n $\approx$ 0.5 cm$^{-3}$) and an overall low mass fraction \changes{to} not produce appreciable \cii\ emission \citep[see also][]{pineda2013, fahrion2017, herrera2017}. Therefore the \cii\ emission associated with the atomic gas is due to CNM. In the equation above $I^{\rm atomic}_{\rm [CII]} \cong I^{\rm CNM}_{\rm [CII]}$.

\cii\ emission is the dominant cooling source in the CNM \citep[]{wolfire1995, Wolfire2003, Draine2011}. The cooling rate per H nucleon is

\begin{equation}
    \Lambda_{\mathrm{[CII]}} = \frac{4 \pi I^{\mathrm{tot}}_{\mathrm{[CII]}}}{N_{\rm HI}}
\label{eq:cool}
\end{equation}
where $N_{\rm HI}$ is the column density of the \hi\ gas in cm$^{-2}$ derived from the 21\,cm spin-flip transition, to which both the WNM and CNM contribute. The fraction of \hi\ column density in the CNM is $f_{\mathrm{CNM}}=N_{\rm HI}^{\rm CNM}/N_{\rm HI}$, with values likely in the range $f_{\mathrm{CNM}}=0.3 - 0.7$ \citep[]{heiles2003}. Therefore,

\begin{equation}
    \Lambda^{\rm CNM}_{\rm [CII]} = \frac{4 \pi I^{\rm CNM}_{\rm [CII]}}{f_{\rm CNM} N_{\rm HI}}.
\label{eq:coolCNM}
\end{equation}

We can then relate the observed cooling rate as defined in \autoref{eq:cool} to the CNM cooling rate using

\begin{equation}
    \Lambda_{\rm [CII]} = \Lambda^{\rm CNM}_{\rm [CII]} \frac{f_{\rm CNM}}{(1-f_{\rm ion})(1-f_{\rm mol})(1-f_{\rm H_2})}.
\end{equation}

With \autoref{eq:craw} as the expression for the \cii\ intensity, we rewrite the cooling rate as

\begin{equation}
    \Lambda^{\rm CNM}_{\rm [CII]} = \frac{2.9 \times 10^{-20} K  N_{\rm C^+}^{\rm CNM}}{f_{\rm CNM}  N_{\rm HI}},
\end{equation}
\noindent where $K$ is 
\begin{equation}
     K = \frac{2e^{-91.2/T}} {1 + 2e^{-91.2/T} + A_{ul}/(\Sigma R_{ul,i} n_i)}.
\label{eq:K}
\end{equation}

Assuming the carbon abundance is the same for the CNM and WNM, and \changes{assuming all gas-phase carbon is} C$^+$, then $ N_{\rm C^+}^{\rm CNM}/({f_{\rm CNM}N_{\rm HI}}) = \mathrm{(C/H)}^{\rm CNM}$, and the final expression for the observed cooling rate is

\begin{equation}
    \Lambda_{\rm [CII]} = \frac{2.9 \times 10^{-20} K \mathrm{(C/H)}^{\rm CNM} f_{\rm CNM}}{(1-f_{\rm ion})(1-f_{\rm mol})(1-f_{\rm H_2})}.
\label{eq:coolfinal}
\end{equation}

\begin{figure*}
\centering
\includegraphics[width=1\textwidth]{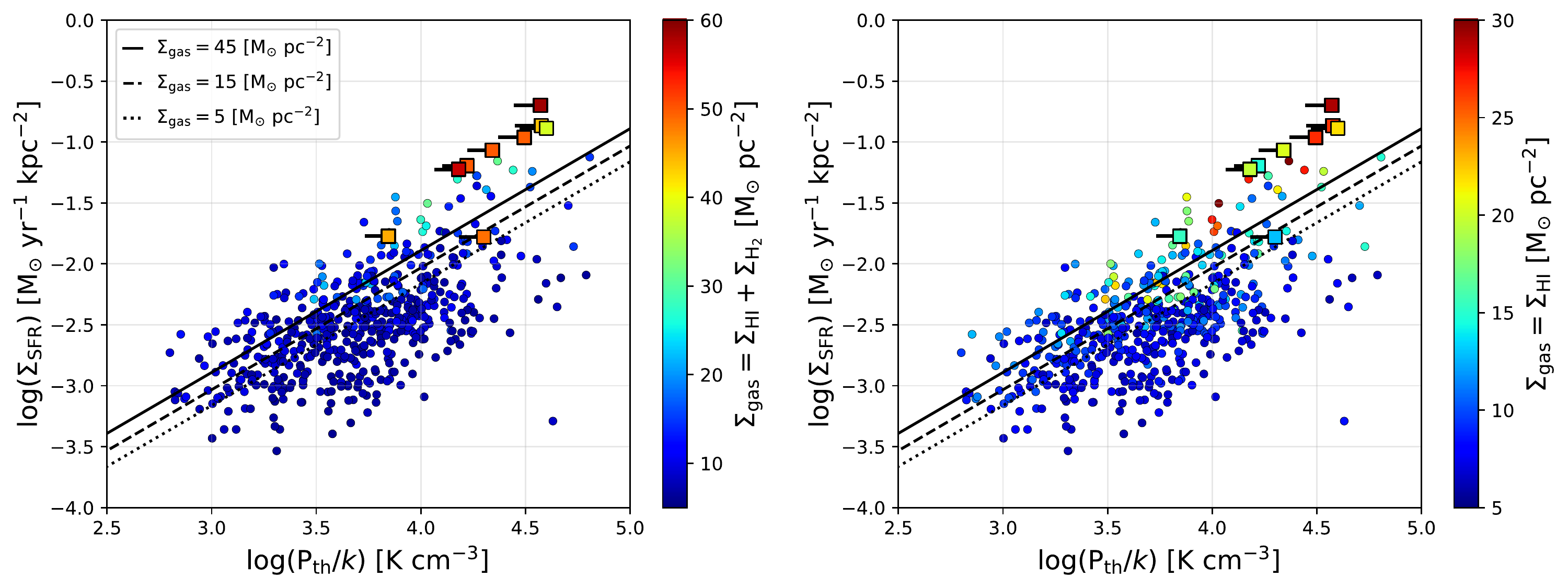}
\caption{The thermal pressure vs. star formation rate where $f_{\rm{ion}} = 0.12$ and $f_{\rm{H_2}} = 0.3$. The KINGFISH data from \citet[]{herrera2017} are circles and the 3$\sigma$ stacked data bins in this study correspond to squares. \changes{The color coding represents the gas surface density, with the total presented ($\rm \Sigma_{gas} = \Sigma_{H{\small I}} + \Sigma_{H_2}$) on the left panel and the \hi\ gas surface density ($\rm \Sigma_{H{\small I}}$) shown on the right.} The horizontal lines to the left of the square points show how the pressure would change if $f_{\rm{ion}} = 0.3$, as is set in the KINGFISH data. \changes{The vertical lines on the plot represent theortical predictions from \citet[]{Wolfire2003} for different values of $\rm \Sigma_{gas}$.} Our estimates follow the same trends as the KINGFISH data but they are localized on the higher pressure region rather than the median, likely due to the bias towards regions of higher \sSFR\ \changes{and $\rm \Sigma_{gas}$.}} 
\label{fig:pressure}
\end{figure*}

We then solve for the density, $n_i$, in  \autoref{eq:K}. The collisional partners with $\rm C^+$ in the CNM will be atomic hydrogen and helium. Assuming the cosmic abundance ratio of 10 to 1 for hydrogen to helium, the sum over collisional partners in \autoref{eq:K} simplifies to $1.038R_{ul}(H^0)n_{\rm CNM}$ \citep{Draine2011}. The thermal pressure of the CNM is 

\begin{equation}
    P_{th} = n_{\rm CNM}T ~ \rm K ~ cm^{-3}
\end{equation}

To calculate $P_{th}$ in the CNM, we assume a temperature of $T_{\rm{CNM}} = 100\,K$ \citep[e.g.,][]{Gerin15}. A variation of $T_{\rm{CNM}}$ of a factor of 2 will alter the thermal pressure by $\sim$30\%. We obtain the gas-phase carbon abundance through the oxygen abundance gradients measured for these galaxies \citep{Pilyugin2014} and convert these oxygen abundances into carbon abundances through the relation used in MAPPINGS \citep[]{Nicholls2017}

\begin{equation}
    \log(\mathrm{C/H}) = \log(\mathrm{O/H}) + \log(10^{-1.00} + 10^{2.72 + \log(\mathrm{O/H})}).
\end{equation}

% specify that the abundance is gas 
\noindent\citet[]{herrera2017} normalize the expression above to recover the local Galactic ISM  gas phase carbon abundance of $\mathrm{C/H} = 1.5 \times 10^{-4}$ \citep[]{Gerin15} with an input oxygen gas phase abundance of 12 + log(O/H) = 8.65 \citep{Simon-Diaz2011}. 
We set the fraction of CNM in the atomic gas to $f_{\rm{CNM}} = 0.5$, which is consistent with the results of \citet[]{heiles2003} and \citet{pineda2013}. 
The fraction of \htwo\ is uncertain but we set it to $f_{\rm{H_2}} = 0.3$, motivated by \citet{pineda2013} who use \cii\ observations of the plane of the Milky Way to find that the CO-dark gas contributes $\sim30\%$ of the total \cii\ emission.  
Our \nii\ SOFIA/GREAT observations have an average upper limit of $f_{\rm ion}=0.12$ for the \cii\ emission from ionized gas (c.f., \autoref{fig:NII}), although we also include the result of setting $f_{\rm{ion}} = 0.3$, as in \citet{herrera2017}.

In addition to the assumptions described above, we use the results in \autoref{tab:decomp_tab} for the value of $f_{\rm mol}$. We then estimate the thermal pressure of the CNM for each stacked bin through the \cii\ cooling rate given in \autoref{eq:coolfinal}. Those results are compared to the calculated thermal pressures found in \citet[]{herrera2017} and shown in \autoref{fig:pressure}.

The thermal pressure of the CNM calculated in this study range from $\log(P_{th}/k) = 3.8 - 4.6 \mathrm{~[K ~ cm^{-3}]}$. These pressures correspond to a density range of 75~$\rm cm^{-3}$ to 400~$\rm cm^{-3}$ when assuming $T_{\rm{CNM}} = \rm 100\,K$. The effect of increasing $f_{\rm{ion}}$ from 0.12 to 0.3, the value \citet{herrera2017} uses \changes{for the KINGFISH sample}, decreases the thermal pressure by 0.1 dex on average.

These pressures are slightly higher than those derived in the atomic disk for the KINGFISH sample, but broadly follow the same trends with \changes{gas surface density} and \sSFR. \citet[]{herrera2017} identify atomic-dominated regions by finding where the surface density of atomic gas is larger than that of molecular gas \changes{and thus are biased towards quiescent regions with lower gas surface densities ($\rm \Sigma_{gas}$). The technique in this work spectrally identifies the \cii\ coming only from the atomic gas and allows us to calculate $\rm P_{th}/k$ for a wider range of regions with higher \sSFR\ and $\rm \Sigma_{gas}$. \autoref{fig:pressure} shows that this work has much larger total gas surface density values due to the inclusion of regions with appreciable molecular gas. The atomic gas surface density values of these regions, however, are comparable to those in \citet[]{herrera2017}. The KINGFISH data and analytic work by \citet[]{Wolfire2003} show that an increase in \sSFR\ leads to higher thermal pressures. At fixed \sSFR , however, larger gas surface densities decrease the thermal pressure, leading to different predicted slopes for the relationship between the \sSFR\ and $\rm P_{th}/k$ as seen in \autoref{fig:pressure}. Because the regions in this work have high \sSFR\ and $\rm \Sigma_{gas}$, these data are on the upper end of the trend between star formation rate and thermal pressure. Similar to \citet[]{herrera2017}, the trends here generally agree with the theory, but there is a wider dispersion in the relationship between \sSFR , $\rm \Sigma_{gas}$, $\rm P_{th}/k$, possibly due to regions continuing to evolve to equilibrium or observational uncertainties in these measurements.}
%mention how this technique allow us to use regions that have a molecular component, have higher SFR, 
%pressure equilibrium is more dynamical than thermal
%there is a general trend that agrees with the theory, but the dispersion is higher than theory predicts 

% All of the property bins presented here have a higher molecular gas surface density, and would not have been included in the \citet{herrera2017} sample. Because of the relationship between CO and SFR \citep{Kennicutt1998}, the regions we use are biased toward higher star formation rates than those of \citet{herrera2017}.

\citet{Gerin15} compute thermal pressures in Milky Way star-forming regions using \cii\ and [CI] far-infrared/sub-mm observations toward bright dust continuum regions on the Galactic plane, and find a median of $\log(P_{th}/k) = 3.8 \mathrm{~K ~ cm^{-3}}$ and a maximum of $\log(P_{th}/k) = 4.3 \mathrm{~K ~ cm^{-3}}$, comparable to our result. \citet{Jenkins2011} use CI ultraviolet absorption measurements toward Milky Way stars within 3\,kpc to estimate a median pressure of $\log(P_{th}/k) = 3.6 \mathrm{~K ~ cm^{-3}}$ with a log-normal distribution, although finding an excess of $\log(P_{th}/k) >4.0 \mathrm{~K ~ cm^{-3}}$ of pressures and a positive correlation with radiation field (and star formation activity). \citet{Goldsmith2018} use velocity resolved \cii\ observations and 21 cm absorption spectra and find a pressure range of $\log(P_{th}/k) = 3.3 - 4.0 \mathrm{~[K ~ cm^{-3}]}$ in atomic gas dominated lines of sight. Also using \cii\ observations, \citet{Velusamy2017} compute the pressure by isolating the \htwo\ gas and find areas of higher pressure, $\log(P_{th}/k) > 4 \mathrm{~K ~ cm^{-3}}$, in high star-forming regimes. 

Outside of the Milky Way, \citet{Welty2016} use CI ultraviolet absorption measurements to identify the CNM thermal pressure in sight lines of Magellanic Clouds and compute pressures that range from $\log(P_{th}/k) = 3.6 - 5.1 \mathrm{~[K ~ cm^{-3}]}$, agreeing well with the range of thermal pressures we find for M101 and NGC~6946. The authors hypothesize that the higher pressures may be due to the enhanced radiation fields in the CNM of the Magellanic Clouds. Energetic feedback from stellar winds, supernova remnants, and star formation may also play a role in increasing the overall thermal pressure of in the Magellanic Clouds. It is possible that the same effects may be contributing to the higher thermal pressures found in M101 and NGC~6946. 

The trends we find with \sSFR\ and \changes{gas surface density} generally agree with the observational work of \citet[]{herrera2017} and the analytic modeling work by \citet[]{Wolfire2003}. The thermal pressure increases with increasing \sSFR, possibly caused by the relation between \sSFR\ and $G_0$, the far-ultraviolet (FUV) intensity field \citep{Dopita1985, ostriker2010}, or through stellar feedback from winds or supernovae \citep{Hayward2017, Barrera2021}.

\subsection{Origins of the [CII] emission}
\label{sec:compare}

Studies that explored the multi-phase nature of \cii\ emission in NGC~6946 did so initially based on spatial information rather than spectral profiles. Using the {\em Kuiper} Airborne Observatory, \citet{Madden1993} found an extended component of \cii\ emission that they attribute to the atomic gas. Following up that study, measurements of NGC~6946 using the Infrared Space Observatory found that $\lessapprox$40\% of the \cii\ emission comes from the diffuse galaxy disk \citep{Contursi2002}. Through PDR modeling, these authors find that the majority of the \cii\ emission associated with the \hi\ gas arises from dense \hi, likely from the photodissociation of \htwo\ on molecular cloud surfaces, and is consistent with the density and pressure computed in our study. More recently, \citet{Bigiel2020} use \cii\ SOFIA/FIFI-LS data of NGC~6946 to find that 73\% of the \cii\ luminosity comes from the spiral arms, 19\% is from the central regions, and 8\% is in the interarm regions. It is difficult to compare directly to this work, because we separate our regions based on star formation rate and metallicity, but these spatial results are broadly consistent with what is measured here.

Previous work studying the origin of \cii\ emission with velocity resolved \cii\ data and some form of profile decomposition are broadly consistent with our conclusions. \citet[]{fahrion2017} use SOFIA/GREAT observations of the dwarf galaxy NGC~4214 and a similar spectral decomposition method to identify the origin of \cii\ emission in five regions at a resolution of $\sim$200 pc. They find on average that 54\% of the \cii\ emission is associated with the CO profiles and 46\% is associated with the \hi\ profiles, in agreement with our finding (see \autoref{tab:decomp_tab}). The authors state, however, that only about 5 - 11\% of the \cii\ emission originates in the CNM because they take a narrower definition of the CNM than we use here. In order to reproduce the \cii\ emission associated with the \hi\ gas in NGC 4214, \citet{fahrion2017} calculate a density of $\sim$1000 cm$^{-3}$ at a temperature of 80 K, concluding there is a denser atomic phase than the classical CNM associated with the \cii\ that has similar broad wings to the \hi\ profile. Our pressures in fact suggest that this ``high pressure CNM phase'' is fairly common in star-forming regions.

\citet{Mookerjea2016} use \textit{Herschel}/HIFI velocity resolved \cii\ data of M33 at 50 pc resolutions and a method of combining the CO and \hi\ profiles to reproduce the \cii\ spectra that is identical to the method presented in this study. In 20 different regions that cover the center of M33 and one of its large \hii\ regions, they find that 8-85\% of the \cii\ emission comes from the atomic gas. They calculate CNM densities that range from 150~$\rm cm^{-3}$ to 1500~$\rm cm^{-3}$, depending on the given region in their sample. With these high densities, they conclude that the \cii\ originating from the majority of the atomic medium comes from the atomic envelopes of molecular PDRs. Therefore, \citet{Mookerjea2016} is consistent with other studies \citep[e.g.][]{Contursi2002, fahrion2017}, including the work in this paper, that associate portions of the \cii\ emission with a dense, high pressure, atomic phase. 

Studies that focus on \cii\ in the Magellanic Clouds have higher spatial resolutions, on the order of a few parsecs compared to the $\sim$500~pc regions in this work. \citet[]{okada2015, okada2019} use spectrally resolved SOFIA/GREAT observations of CO and \cii\ in the LMC at spatial resolutions of $\sim$4~pc to show that the CO spectra alone cannot explain 30\% - 60\% of the \cii\ emission in the LMC, suggesting the presence of CO-dark gas. By matching the wide wings in the \cii\ spectra with the \hi\ profile, they find that less than 15\% of the \cii\ emission comes from the atomic phase on average. Other studies in the SMC that use velocity resolved \cii\ data also observe a small contribution from the atomic gas to the overall \cii\ emission \citep{requena2016}. \citet[]{Lebouteiller2019} use SOFIA/GREAT data of the LMC and employ a Bayesian approach to decompose the line profiles of each tracer into multiple components per region, which are analyzed individually. In order to find the contribution from atomic gas, they estimate the density using the \hi\ column density and the average cloud sizes in the LMC found by \citet[]{Indebetouw2013}, computing densities that range between a few cm$^{-3}$ to 10$^3$ cm$^{-3}$. With this method, the atomic phase contributes about 30\% of the total \cii\ emission in regions with faint \cii\ emission. In bright regions, CO-dark gas associated with \cii\ dominates, contributing 95\% to the emission. 

In this study we find that the atomic gas contributes $\sim$50\% or more to the overall \cii\ emission, similar to the other spectral decomposition studies on scales between 50 and 200~pc \citep{Mookerjea2016, fahrion2017}. At these resolutions, multiple \hii\ regions, PDR complexes, and extended gas are averaged into one beam. It is likely that the discrepancy of the importance of the contribution from the atomic phase between the Magellanic Clouds (where atomic gas contributes less to the \cii\ emission) and other studies is caused by the difference in spatial scales, as we would expect the more extended components to contribute more on the larger scales. In the literature, as well as in our study, there is a tendency to find that the contribution of the atomic gas to the \cii\ emission increases in regions with fainter \cii\ emission \citep{fahrion2017, Lebouteiller2019}. Neither the literature nor this study finds that the origin of \cii\ emission has a consistent dependence on the star formation rate, metallicity, \changes{or galactocentric radius} of the region.

\section{Summary \& Conclusions}\label{sec:conclusions}
We present two cycles of SOFIA/GREAT velocity resolved 158\,\micron\ \cii\ and 205\,\micron\ \nii\ observations of the nearby galaxies M101 and NGC 6946. These observations have a spatial resolution of $\sim$500\,pc and probe a variety of regions that range in star formation rate and metallicity. We compare the velocity resolved \cii\ spectra to ancillary \hi\ spectra from the THINGS survey and CO spectra from the HERACLES survey. The goal of this study is to determine the origin of the multi-phase \cii\ emission through spectral decomposition using \hi\ as a tracer of atomic gas and CO as a tracer of molecular gas. We model the \cii\ emission as a linear combination of the \hi\ and CO spectra and identify the fraction of the \cii\ emission associated with each phase. After isolating only the \cii\ emission coming from the atomic phase, we compute the cooling rate per hydrogen nucleus (\autoref{eq:coolfinal}) in order to solve for the thermal pressure of the CNM. Our main results are as follows: 

\begin{enumerate}
\itemsep-0.5em
    \item We find that the \hi\ spectral profiles are on average 29\% wider than the \cii\ spectra, while the CO spectra are 25\% narrower than the \cii\ spectra (\autoref{fig:gau}). The \cii\ linewidths lie in between the CO and HI \citep[see also][]{deBlok2016, requena2016, Lebouteiller2019}, suggesting that the \cii\ originates from both molecular and atomic gas. 

    \item We find that the neutral gas (atomic and molecular) contributes at least 88\% to the \cii\ emission on the average of our pointings, based on the \nii\ 205\,$\mu$m upper limit data acquired by GREAT (\autoref{fig:NII}). This agrees with other studies that use \nii\ observations to model the ionized gas contribution \citep{pineda2013, Croxall2013, Lebouteiller2019}. Thus the ionized gas has a negligible contribution to the \cii\ emission.

    \item To quantify the reliability and uniqueness of our \cii\ decomposition methodology we use template spectra derived from our data to run a series of simulations. We find that a peak SNR$\sim10-15$ is required to accurately decompose the \cii\ emission into the atomic and molecular components using our linear combination methodology (\S\ref{sec:eval}, Figures \ref{fig:accuracy} and \ref{fig:unique}). 
    
    \item We perform our analysis on spectra stacked in bins of \sSFR, metallicity, \changes{and normalized galactocentric radius} for different samples based on SNR and intensity. We find that over all the spectra the atomic phase contributes $\gtrsim$50\% or more to the the \cii\ emission  ($f_{\rm mol}\simeq48\%$, $f_{\rm atomic}\simeq52\%$ when stacking all $\geq3\sigma$ spectra), with a weak but consistent trend for the fainter \cii\ emission to have an increasing contribution from the atomic medium (\autoref{tab:decomp_tab}, \autoref{fig:decomp_sum}). We also perform our decomposition on the three individual spectra with sufficient SNR to produce meaningful results, and confirm this finding that on average $45\%-55\%$ of emission arises from molecular and atomic gas, respectively (\autoref{tab:tab_ind}).
    
    \item While the fraction of atomic or molecular gas associated with the \cii\ emission has no clear dependence with \sSFR, metallicity \changes{or galactocentric radius}, there is a significant difference in the results of the \cii\ decomposition when comparing spectra in M101 and NGC~6946. The \cii\ pointings in M101 are more dominated by the atomic gas ($\fmol\simeq0.28\pm0.09$) while those in NGC~6946 appear more associated with the molecular gas ($\fmol\simeq0.59\pm0.10$).

    \item At the lowest metallicities probed and in the faintest \cii\ spectra of M101 there is a tentative hint of an extra component in the \cii\ emission which may be associated with a ``CO-dark'' phase (\autoref{fig:lowZ}). Evidence of \cii\ tracing CO-dark gas is also seen in a variety of other \cii\ studies, especially in regions with low metallicities \citep[e.g.][]{pineda2013, fahrion2017, Lebouteiller2019}.

    \item From the \cii\ emission, we find a thermal pressure of $\log(P_{th}/k) = 3.8 - 4.6 \mathrm{~[K ~ cm^{-3}]}$ in the atomic gas in our pointings (\autoref{fig:pressure}). This is somewhat higher than other estimates of the thermal pressure in the atomic phase \citep{Jenkins2011, Gerin15, herrera2017}. We suspect this is likely due to the comparatively high \sSFR\ in our regions. Other studies of the origin of \cii\ also report a significant contribution from atomic gas that is at higher densities and pressures than the traditional Cold Neutral Medium \citep{Contursi2002, Mookerjea2016, fahrion2017}.

\end{enumerate}

Because SNR $\gtrsim$ 10-15 are needed for a meaningful decomposition of the velocity resolved \cii\ emission, we highlight the importance of acquiring high SNR observations of extragalactic \cii, which can be time-consuming. As demonstrated in this work, however, velocity resolved \cii\ observations that enable kinematic decomposition of the emission are a useful tool for understanding the origin of \cii\ and the physical conditions in the ISM.

\acknowledgments
\changes{We thank the referee for comments and suggestions that greatly improved the paper.} ET would like to thank Laura Lenki\'c and Ramsey Karim for thoughtful discussions and Richard Cosentino for helpful comments on the draft manuscript. This paper is based on observations made with the NASA/DLR Stratospheric Observatory for Infrared Astronomy (SOFIA). GREAT/upGREAT, the German Receiver for Astronomy at Terahertz Frequencies, was developed and built by a consortium of German research institutes (MPI for Radio Astronomy/MPIfR, Bonn and KOSMA/Cologne University, in collaboration with the DLR Institute for Planetary Research, Berlin, and the MPI for Solar System Research, Göttingen). The development of the instrument was financed by the participating institutes, the Max Planck Society the Deutsche Forschungsgemeinschaft, and DLR. SOFIA is jointly operated by the Universities Space Research Association, Inc. (USRA), under NASA contract NNA17BF53C, and the Deutsches SOFIA Institut (DSI) under DLR contract 50 OK 0901 to the University of Stuttgart. Financial support for this work was provided by NASA through awards USRA SOFIA 02\_0098, 04\_0151, 07\_0126 issued by USRA. This work made use of HERACLES, The HERA CO-Line Extragalactic Survey \citep{leroy2007}, and THINGS, The HI Nearby Galaxy Survey \citep{Walter2008}. 

\software{Astropy \citep{astropy}, 
MatPlotLib \citep{matplotlib}, 
NumPy \citep{numpy},  
SciPy \citep{scipy}, 
CLASS/GILDAS \citep{2005sf2a.conf..721P,2013ascl.soft05010G}\footnote{\url{https://www.iram.fr/IRAMFR/GILDAS/}}, 
Spectral-Cube \citep{2019zndo...2573901G}\footnote{\url{https://spectral-cube.readthedocs.io/en/latest/}}}

\bibliographystyle{apj}
\bibliography{references.bib}

\begin{thebibliography}{}
\expandafter\ifx\csname natexlab\endcsname\relax\def\natexlab#1{#1}\fi

\bibitem[{{Accurso} {et~al.}(2017){Accurso}, {Saintonge}, {Catinella},
  {Cortese}, {Dav{\'e}}, {Dunsheath}, {Genzel}, {Gracia-Carpio}, {Heckman},
  {Jimmy}, {Kramer}, {Li}, {Lutz}, {Schiminovich}, {Schuster}, {Sternberg},
  {Sturm}, {Tacconi}, {Tran}, \& {Wang}}]{accurso2017}
{Accurso}, G., {Saintonge}, A., {Catinella}, B., {et~al.} 2017, \mnras, 470,
  4750

\bibitem[{{Anand} {et~al.}(2018){Anand}, {Rizzi}, \& {Tully}}]{Anand2018}
{Anand}, G.~S., {Rizzi}, L., \& {Tully}, R.~B. 2018, \aj, 156, 105

\bibitem[{Aniano {et~al.}(2011)Aniano, Draine, Gordon, \&
  Sandstrom}]{Aniano2011}
Aniano, G., Draine, B.~T., Gordon, K.~D., \& Sandstrom, K. 2011, Publications
  of the Astronomical Society of the Pacific, 123, 1218–1236

\bibitem[{{Aniano} {et~al.}(2020){Aniano}, {Draine}, {Hunt}, {Sandstrom},
  {Calzetti}, {Kennicutt}, {Dale}, {Galametz}, {Gordon}, {Leroy}, {Smith},
  {Roussel}, {Sauvage}, {Walter}, {Armus}, {Bolatto}, {Boquien}, {Crocker}, {De
  Looze}, {Donovan Meyer}, {Helou}, {Hinz}, {Johnson}, {Koda}, {Miller},
  {Montiel}, {Murphy}, {Rela{\~n}o}, {Rix}, {Schinnerer}, {Skibba}, {Wolfire},
  \& {Engelbracht}}]{Aniano2020}
{Aniano}, G., {Draine}, B.~T., {Hunt}, L.~K., {et~al.} 2020, \apj, 889, 150

\bibitem[{{Barrera-Ballesteros} {et~al.}(2021){Barrera-Ballesteros},
  {S{\'a}nchez}, {Heckman}, {Wong}, {Bolatto}, {Ostriker}, {Rosolowsky},
  {Carigi}, {Vogel}, {Levy}, {Colombo}, {Luo}, {Cao}, \& {the EDGE-CALIFA
  team}}]{Barrera2021}
{Barrera-Ballesteros}, J.~K., {S{\'a}nchez}, S.~F., {Heckman}, T., {et~al.}
  2021, arXiv e-prints, arXiv:2101.04683

\bibitem[{{Barrett} {et~al.}(2005){Barrett}, {Hunter}, {Miller}, {Hsu}, \&
  {Greenfield}}]{matplotlib}
{Barrett}, P., {Hunter}, J., {Miller}, J.~T., {Hsu}, J.-C., \& {Greenfield}, P.
  2005, in Astronomical Society of the Pacific Conference Series, Vol. 347,
  Astronomical Data Analysis Software and Systems XIV, ed. P.~{Shopbell},
  M.~{Britton}, \& R.~{Ebert}, 91

\bibitem[{{Bennett} {et~al.}(1994){Bennett}, {Fixsen}, {Hinshaw}, {Mather},
  {Moseley}, {Wright}, {Eplee}, {Gales}, {Hewagama}, {Isaacman}, {Shafer}, \&
  {Turpie}}]{Bennett1994}
{Bennett}, C.~L., {Fixsen}, D.~J., {Hinshaw}, G., {et~al.} 1994, \apj, 434, 587

\bibitem[{{Bigiel} {et~al.}(2020){Bigiel}, {de Looze}, {Krabbe}, {Cormier},
  {Barnes}, {Fischer}, {Bolatto}, {Bryant}, {Colditz}, {Geis}, {Herrera-Camus},
  {Iserlohe}, {Klein}, {Leroy}, {Linz}, {Looney}, {Madden}, {Poglitsch},
  {Stutzki}, \& {Vacca}}]{Bigiel2020}
{Bigiel}, F., {de Looze}, I., {Krabbe}, A., {et~al.} 2020, arXiv e-prints,
  arXiv:2011.02498

\bibitem[{{Boselli} {et~al.}(2002){Boselli}, {Gavazzi}, {Lequeux}, \&
  {Pierini}}]{Boselli2002}
{Boselli}, A., {Gavazzi}, G., {Lequeux}, J., \& {Pierini}, D. 2002, \aap, 385,
  454

\bibitem[{{Braun} \& {Walterbos}(1985)}]{braun1985}
{Braun}, R., \& {Walterbos}, R.~A.~M. 1985, \aap, 143, 307

\bibitem[{{Calzetti} {et~al.}(2007){Calzetti}, {Kennicutt}, {Engelbracht},
  {Leitherer}, {Draine}, {Kewley}, {Moustakas}, {Sosey}, {Dale}, {Gordon},
  {Helou}, {Hollenbach}, {Armus}, {Bendo}, {Bot}, {Buckalew}, {Jarrett}, {Li},
  {Meyer}, {Murphy}, {Prescott}, {Regan}, {Rieke}, {Roussel}, {Sheth}, {Smith},
  {Thornley}, \& {Walter}}]{calzetti2007}
{Calzetti}, D., {Kennicutt}, R.~C., {Engelbracht}, C.~W., {et~al.} 2007, \apj,
  666, 870

\bibitem[{{Contursi} {et~al.}(2002){Contursi}, {Kaufman}, {Helou},
  {Hollenbach}, {Brauher}, {Stacey}, {Dale}, {Malhotra}, {Rubio}, {Rubin}, \&
  {Lord}}]{Contursi2002}
{Contursi}, A., {Kaufman}, M.~J., {Helou}, G., {et~al.} 2002, \aj, 124, 751

\bibitem[{{Cormier} {et~al.}(2015){Cormier}, {Madden}, {Lebouteiller}, {Abel},
  {Hony}, {Galliano}, {R{\'e}my-Ruyer}, {Bigiel}, {Baes}, {Boselli},
  {Chevance}, {Cooray}, {De Looze}, {Doublier}, {Galametz}, {Hughes},
  {Karczewski}, {Lee}, {Lu}, \& {Spinoglio}}]{cormier2015}
{Cormier}, D., {Madden}, S.~C., {Lebouteiller}, V., {et~al.} 2015, \aap, 578,
  A53

\bibitem[{{Cormier} {et~al.}(2019){Cormier}, {Abel}, {Hony}, {Lebouteiller},
  {Madden}, {Polles}, {Galliano}, {De Looze}, {Galametz}, \&
  {Lambert-Huyghe}}]{cormier2019}
{Cormier}, D., {Abel}, N.~P., {Hony}, S., {et~al.} 2019, \aap, 626, A23

\bibitem[{{Crawford} {et~al.}(1985){Crawford}, {Genzel}, {Townes}, \&
  {Watson}}]{crawford1985}
{Crawford}, M.~K., {Genzel}, R., {Townes}, C.~H., \& {Watson}, D.~M. 1985,
  \apj, 291, 755

\bibitem[{{Croxall} {et~al.}(2013){Croxall}, {Smith}, {Brandl}, {Groves},
  {Kennicutt}, {Kreckel}, {Johnson}, {Pellegrini}, {Sandstrom}, {Walter},
  {Armus}, {Beir{\~a}o}, {Calzetti}, {Dale}, {Galametz}, {Hinz}, {Hunt},
  {Koda}, \& {Schinnerer}}]{Croxall2013}
{Croxall}, K.~V., {Smith}, J.~D., {Brandl}, B.~R., {et~al.} 2013, \apj, 777, 96

\bibitem[{{Croxall} {et~al.}(2017){Croxall}, {Smith}, {Pellegrini}, {Groves},
  {Bolatto}, {Herrera-Camus}, {Sand strom}, {Draine}, {Wolfire}, {Armus},
  {Boquien}, {Brandl}, {Dale}, {Galametz}, {Hunt}, {Kennicutt}, {Kreckel},
  {Rigopoulou}, {van der Werf}, \& {Wilson}}]{Croxall2017}
{Croxall}, K.~V., {Smith}, J.~D., {Pellegrini}, E., {et~al.} 2017, \apj, 845,
  96

\bibitem[{{de Blok} {et~al.}(2016){de Blok}, {Walter}, {Smith},
  {Herrera-Camus}, {Bolatto}, {Requena-Torres}, {Crocker}, {Croxall},
  {Kennicutt}, {Koda}, {Armus}, {Boquien}, {Dale}, {Kreckel}, \&
  {Meidt}}]{deBlok2016}
{de Blok}, W.~J.~G., {Walter}, F., {Smith}, J. D.~T., {et~al.} 2016, \aj, 152,
  51

\bibitem[{{De Looze} {et~al.}(2014){De Looze}, {Cormier}, {Lebouteiller},
  {Madden}, {Baes}, {Bendo}, {Boquien}, {Boselli}, {Clements}, {Cortese},
  {Cooray}, {Galametz}, {Galliano}, {Graci{\'a}-Carpio}, {Isaak}, {Karczewski},
  {Parkin}, {Pellegrini}, {R{\'e}my-Ruyer}, {Spinoglio}, {Smith}, \&
  {Sturm}}]{DeLooze2014}
{De Looze}, I., {Cormier}, D., {Lebouteiller}, V., {et~al.} 2014, \aap, 568,
  A62

\bibitem[{{Dopita}(1985)}]{Dopita1985}
{Dopita}, M.~A. 1985, \apjl, 295, L5

\bibitem[{{Draine}(2011)}]{Draine2011}
{Draine}, B.~T. 2011, {Physics of the Interstellar and Intergalactic Medium}
  (Princeton University Press)

\bibitem[{{Draine} {et~al.}(2007){Draine}, {Dale}, {Bendo}, {Gordon}, {Smith},
  {Armus}, {Engelbracht}, {Helou}, {Kennicutt}, {Li}, {Roussel}, {Walter},
  {Calzetti}, {Moustakas}, {Murphy}, {Rieke}, {Bot}, {Hollenbach}, {Sheth}, \&
  {Teplitz}}]{Draine2007}
{Draine}, B.~T., {Dale}, D.~A., {Bendo}, G., {et~al.} 2007, \apj, 663, 866

\bibitem[{{Fahrion} {et~al.}(2017){Fahrion}, {Cormier}, {Bigiel}, {Hony},
  {Abel}, {Cigan}, {Csengeri}, {Graf}, {Lebouteiller}, {Madden}, {Wu}, \&
  {Young}}]{fahrion2017}
{Fahrion}, K., {Cormier}, D., {Bigiel}, F., {et~al.} 2017, \aap, 599, A9

\bibitem[{{Fern{\'a}ndez Arenas} {et~al.}(2018){Fern{\'a}ndez Arenas},
  {Terlevich}, {Terlevich}, {Melnick}, {Ch{\'a}vez}, {Bresolin}, {Telles},
  {Plionis}, \& {Basilakos}}]{Fernandez2018}
{Fern{\'a}ndez Arenas}, D., {Terlevich}, E., {Terlevich}, R., {et~al.} 2018,
  \mnras, 474, 1250

\bibitem[{{Field} {et~al.}(1969){Field}, {Goldsmith}, \& {Habing}}]{field1969}
{Field}, G.~B., {Goldsmith}, D.~W., \& {Habing}, H.~J. 1969, \apjl, 155, L149

\bibitem[{{Geis} \& {Lutz}(2010)}]{Geis2010}
{Geis}, N., \& {Lutz}, D. 2010, {Herschel/PACS Modelled Point-Spread
  Functions}, Tech. rep., {PACS Instrument Control Centre PICC-ME-TN-029}

\bibitem[{{Gerin} {et~al.}(2015){Gerin}, {Ruaud}, {Goicoechea}, {Gusdorf},
  {Godard}, {de Luca}, {Falgarone}, {Goldsmith}, {Lis}, {Menten}, {Neufeld},
  {Phillips}, \& {Liszt}}]{Gerin15}
{Gerin}, M., {Ruaud}, M., {Goicoechea}, J.~R., {et~al.} 2015, \aap, 573, A30

\bibitem[{{Gildas Team}(2013)}]{2013ascl.soft05010G}
{Gildas Team}. 2013, {GILDAS: Grenoble Image and Line Data Analysis Software},
  ascl:1305.010

\bibitem[{{Ginsburg} {et~al.}(2019){Ginsburg}, {Koch}, {Robitaille},
  {Beaumont}, {ZuHone}, {Sipocz}, {Patra}, {Jones}, {Lim}, {Rosolowsky},
  {Stern}, {Earl}, {de Val-Borro}, {jrobbfed}, {shuokong}, {Kepley}, {Sokolov},
  {Badger}, {Maret}, {Garrido}, {Booker}, \& {Tollerud}}]{2019zndo...2573901G}
{Ginsburg}, A., {Koch}, E., {Robitaille}, T., {et~al.} 2019,
  {radio-astro-tools/spectral-cube}, doi:10.5281/zenodo.2573901

\bibitem[{{Glover} \& {Clark}(2012)}]{Glover2012}
{Glover}, S. C.~O., \& {Clark}, P.~C. 2012, \mnras, 421, 9

\bibitem[{{Goldsmith} {et~al.}(2012){Goldsmith}, {Langer}, {Pineda}, \&
  {Velusamy}}]{goldsmith2012}
{Goldsmith}, P.~F., {Langer}, W.~D., {Pineda}, J.~L., \& {Velusamy}, T. 2012,
  \apjs, 203, 13

\bibitem[{{Goldsmith} {et~al.}(2018){Goldsmith}, {Pineda}, {Neufeld},
  {Wolfire}, {Risacher}, \& {Simon}}]{Goldsmith2018}
{Goldsmith}, P.~F., {Pineda}, J.~L., {Neufeld}, D.~A., {et~al.} 2018, \apj,
  856, 96

\bibitem[{{Grenier} {et~al.}(2005){Grenier}, {Casandjian}, \&
  {Terrier}}]{grenier2005}
{Grenier}, I.~A., {Casandjian}, J.-M., \& {Terrier}, R. 2005, Science, 307,
  1292

\bibitem[{{Guan} {et~al.}(2012){Guan}, {Stutzki}, {Graf}, {G{\"u}sten},
  {Okada}, {Requena-Torres}, {Simon}, \& {Wiesemeyer}}]{Guan2012}
{Guan}, X., {Stutzki}, J., {Graf}, U.~U., {et~al.} 2012, \aap, 542, L4

\bibitem[{{Harris} {et~al.}(2020){Harris}, {Millman}, {van der Walt},
  {Gommers}, {Virtanen}, {Cournapeau}, {Wieser}, {Taylor}, {Berg}, {Smith},
  {Kern}, {Picus}, {Hoyer}, {van Kerkwijk}, {Brett}, {Haldane}, {Fern\'{a}ndez
  del R\'{i}o}, {Wiebe}, {Peterson}, {G\'{e}rard-Marchant}, {Sheppard},
  {Reddy}, {Weckesser}, {Abbasi}, {Gohlke}, \& {Oliphant}}]{numpy}
{Harris}, C.~R., {Millman}, K.~J., {van der Walt}, S.~J., {et~al.} 2020, \nat,
  585, 357

\bibitem[{{Hayward} \& {Hopkins}(2017)}]{Hayward2017}
{Hayward}, C.~C., \& {Hopkins}, P.~F. 2017, \mnras, 465, 1682

\bibitem[{{Heiles}(1994)}]{Heiles1994}
{Heiles}, C. 1994, \apj, 436, 720

\bibitem[{{Heiles} \& {Troland}(2003)}]{heiles2003}
{Heiles}, C., \& {Troland}, T.~H. 2003, \apj, 586, 1067

\bibitem[{{Herrera-Camus} {et~al.}(2015){Herrera-Camus}, {Bolatto}, {Wolfire},
  {Smith}, {Croxall}, {Kennicutt}, {Calzetti}, {Helou}, {Walter}, {Leroy},
  {Draine}, {Brandl}, {Armus}, {Sand strom}, {Dale}, {Aniano}, {Meidt},
  {Boquien}, {Hunt}, {Galametz}, {Tabatabaei}, {Murphy}, {Appleton}, {Roussel},
  {Engelbracht}, \& {Beirao}}]{Herrera2015}
{Herrera-Camus}, R., {Bolatto}, A.~D., {Wolfire}, M.~G., {et~al.} 2015, \apj,
  800, 1

\bibitem[{{Herrera-Camus} {et~al.}(2017){Herrera-Camus}, {Bolatto}, {Wolfire},
  {Ostriker}, {Draine}, {Leroy}, {Sandstrom}, {Hunt}, {Kennicutt}, {Calzetti},
  {Smith}, {Croxall}, {Galametz}, {de Looze}, {Dale}, {Crocker}, \&
  {Groves}}]{herrera2017}
{Herrera-Camus}, R., {Bolatto}, A., {Wolfire}, M., {et~al.} 2017, \apj, 835,
  201

\bibitem[{{Herrera-Camus} {et~al.}(2018){Herrera-Camus}, {Sturm},
  {Graci{\'a}-Carpio}, {Lutz}, {Contursi}, {Veilleux}, {Fischer},
  {Gonz{\'a}lez-Alfonso}, {Poglitsch}, {Tacconi}, {Genzel}, {Maiolino},
  {Sternberg}, {Davies}, \& {Verma}}]{Herrera2018}
{Herrera-Camus}, R., {Sturm}, E., {Graci{\'a}-Carpio}, J., {et~al.} 2018, \apj,
  861, 95

\bibitem[{{Heyminck} {et~al.}(2012){Heyminck}, {Graf}, {G{\"u}sten}, {Stutzki},
  {H{\"u}bers}, \& {Hartogh}}]{Heyminck2012}
{Heyminck}, S., {Graf}, U.~U., {G{\"u}sten}, R., {et~al.} 2012, \aap, 542, L1

\bibitem[{{Hollenbach} \& {Tielens}(1999)}]{hollenbach1999}
{Hollenbach}, D.~J., \& {Tielens}, A.~G.~G.~M. 1999, Reviews of Modern Physics,
  71, 173

\bibitem[{{Hoopes} {et~al.}(2001){Hoopes}, {Walterbos}, \&
  {Bothun}}]{Hoopes2001}
{Hoopes}, C.~G., {Walterbos}, R. A.~M., \& {Bothun}, G.~D. 2001, \apj, 559, 878

\bibitem[{{Indebetouw} {et~al.}(2013){Indebetouw}, {Brogan}, {Chen}, {Leroy},
  {Johnson}, {Muller}, {Madden}, {Cormier}, {Galliano}, {Hughes}, {Hunter},
  {Kawamura}, {Kepley}, {Lebouteiller}, {Meixner}, {Oliveira}, {Onishi}, \&
  {Vasyunina}}]{Indebetouw2013}
{Indebetouw}, R., {Brogan}, C., {Chen}, C. H.~R., {et~al.} 2013, \apj, 774, 73

\bibitem[{{Jameson} {et~al.}(2018){Jameson}, {Bolatto}, {Wolfire}, {Warren},
  {Herrera-Camus}, {Croxall}, {Pellegrini}, {Smith}, {Rubio}, {Indebetouw},
  {Israel}, {Meixner}, {Roman-Duval}, {van Loon}, {Muller}, {Verdugo},
  {Zinnecker}, \& {Okada}}]{Jameson2018}
{Jameson}, K.~E., {Bolatto}, A.~D., {Wolfire}, M., {et~al.} 2018, \apj, 853,
  111

\bibitem[{{Jenkins} \& {Tripp}(2011)}]{Jenkins2011}
{Jenkins}, E.~B., \& {Tripp}, T.~M. 2011, \apj, 734, 65

\bibitem[{Jones {et~al.}(2001--)Jones, Oliphant, Peterson, {et~al.}}]{scipy}
Jones, E., Oliphant, T., Peterson, P., {et~al.} 2001--, {SciPy}: Open source
  scientific tools for {Python}, [Online; accessed <today>]

\bibitem[{{Kennicutt} {et~al.}(2008){Kennicutt}, {Lee}, {Funes}, {J.}, {Sakai},
  \& {Akiyama}}]{Kennicutt2008}
{Kennicutt}, Robert~C., J., {Lee}, J.~C., {Funes}, J.~G., {et~al.} 2008, \apjs,
  178, 247

\bibitem[{{Kennicutt} {et~al.}(2003){Kennicutt}, {Armus}, {Bendo}, {Calzetti},
  {Dale}, {Draine}, {Engelbracht}, {Gordon}, {Grauer}, {Helou}, {Hollenbach},
  {Jarrett}, {Kewley}, {Leitherer}, {Li}, {Malhotra}, {Regan}, {Rieke},
  {Rieke}, {Roussel}, {Smith}, {Thornley}, \& {Walter}}]{Kennicutt2003}
{Kennicutt}, Robert~C., J., {Armus}, L., {Bendo}, G., {et~al.} 2003, \pasp,
  115, 928

\bibitem[{{Kennicutt} {et~al.}(2009){Kennicutt}, {Hao}, {Calzetti},
  {Moustakas}, {Dale}, {Bendo}, {Engelbracht}, {Johnson}, \&
  {Lee}}]{Kennicutt2009}
{Kennicutt}, Robert~C., J., {Hao}, C.-N., {Calzetti}, D., {et~al.} 2009, \apj,
  703, 1672

\bibitem[{{Kennicutt} {et~al.}(2011){Kennicutt}, {Calzetti}, {Aniano},
  {Appleton}, {Armus}, {Beir{\~a}o}, {Bolatto}, {Brandl}, {Crocker}, {Croxall},
  {Dale}, {Donovan Meyer}, {Draine}, {Engelbracht}, {Galametz}, {Gordon},
  {Groves}, {Hao}, {Helou}, {Hinz}, {Hunt}, {Johnson}, {Koda}, {Krause},
  {Leroy}, {Li}, {Meidt}, {Montiel}, {Murphy}, {Rahman}, {Rix}, {Roussel},
  {Sandstrom}, {Sauvage}, {Schinnerer}, {Skibba}, {Smith}, {Srinivasan},
  {Vigroux}, {Walter}, {Wilson}, {Wolfire}, \& {Zibetti}}]{Kennicutt2011}
{Kennicutt}, R.~C., {Calzetti}, D., {Aniano}, G., {et~al.} 2011, \pasp, 123,
  1347

\bibitem[{{Kim} {et~al.}(2011){Kim}, {Kim}, \& {Ostriker}}]{Kim2011}
{Kim}, C.-G., {Kim}, W.-T., \& {Ostriker}, E.~C. 2011, \apj, 743, 25

\bibitem[{{Kim} \& {Reach}(2002)}]{Kim2002}
{Kim}, S., \& {Reach}, W.~T. 2002, \apj, 571, 288

\bibitem[{{Kramer} {et~al.}(2013){Kramer}, {Abreu-Vicente},
  {Garc{\'\i}a-Burillo}, {Rela{\~n}o}, {Aalto}, {Boquien}, {Braine},
  {Buchbender}, {Gratier}, {Israel}, {Nikola}, {R{\"o}llig}, {Verley}, {van der
  Werf}, \& {Xilouris}}]{Kramer2013}
{Kramer}, C., {Abreu-Vicente}, J., {Garc{\'\i}a-Burillo}, S., {et~al.} 2013,
  \aap, 553, A114

\bibitem[{{Kreckel} {et~al.}(2020){Kreckel}, {Ho}, {Blanc}, {Glover}, {Groves},
  {Rosolowsky}, {Bigiel}, {Boqu{\'\i}en}, {Chevance}, {Dale}, {Deger},
  {Emsellem}, {Grasha}, {Kim}, {Klessen}, {Kruijssen}, {Lee}, {Leroy}, {Liu},
  {McElroy}, {Meidt}, {Pessa}, {Sanchez-Blazquez}, {Sandstrom}, {Santoro},
  {Scheuermann}, {Schinnerer}, {Schruba}, {Utomo}, {Watkins}, \&
  {Williams}}]{Kreckel2020}
{Kreckel}, K., {Ho}, I.~T., {Blanc}, G.~A., {et~al.} 2020, \mnras, 499, 193

\bibitem[{{Krumholz}(2012)}]{Krumholz2012}
{Krumholz}, M.~R. 2012, \apj, 759, 9

\bibitem[{{Kulkarni} \& {Heiles}(1987)}]{kulkarni1987}
{Kulkarni}, S.~R., \& {Heiles}, C. 1987, {The Atomic Component}, Vol. 134 (D.
  J. Hollenbach \& H. A. Thronson, Jr.), 87

\bibitem[{{Langer} {et~al.}(2014){Langer}, {Velusamy}, {Pineda}, {Willacy}, \&
  {Goldsmith}}]{Langer2014}
{Langer}, W.~D., {Velusamy}, T., {Pineda}, J.~L., {Willacy}, K., \&
  {Goldsmith}, P.~F. 2014, \aap, 561, A122

\bibitem[{{Lebouteiller} {et~al.}(2019){Lebouteiller}, {Cormier}, {Madden},
  {Galametz}, {Hony}, {Galliano}, {Chevance}, {Lee}, {Braine}, {Polles},
  {Reque{\~n}a-Torres}, {Indebetouw}, {Hughes}, \& {Abel}}]{Lebouteiller2019}
{Lebouteiller}, V., {Cormier}, D., {Madden}, S.~C., {et~al.} 2019, \aap, 632,
  A106

\bibitem[{{Leroy} {et~al.}(2007){Leroy}, {Bolatto}, {Stanimirovic}, {Mizuno},
  {Israel}, \& {Bot}}]{leroy2007}
{Leroy}, A., {Bolatto}, A., {Stanimirovic}, S., {et~al.} 2007, \apj, 658, 1027

\bibitem[{{Leroy} {et~al.}(2009){Leroy}, {Walter}, {Bigiel}, {Usero}, {Weiss},
  {Brinks}, {de Blok}, {Kennicutt}, {Schuster}, {Kramer}, {Wiesemeyer}, \&
  {Roussel}}]{Leroy2009}
{Leroy}, A.~K., {Walter}, F., {Bigiel}, F., {et~al.} 2009, \aj, 137, 4670

\bibitem[{{Leroy} {et~al.}(2012){Leroy}, {Bigiel}, {de Blok}, {Boissier},
  {Bolatto}, {Brinks}, {Madore}, {Munoz-Mateos}, {Murphy}, {Sandstrom},
  {Schruba}, \& {Walter}}]{Leroy2012}
{Leroy}, A.~K., {Bigiel}, F., {de Blok}, W.~J.~G., {et~al.} 2012, \aj, 144, 3

\bibitem[{{Madden} {et~al.}(1993){Madden}, {Geis}, {Genzel}, {Herrmann},
  {Jackson}, {Poglitsch}, {Stacey}, \& {Townes}}]{Madden1993}
{Madden}, S.~C., {Geis}, N., {Genzel}, R., {et~al.} 1993, \apj, 407, 579

\bibitem[{{Madden} {et~al.}(1997){Madden}, {Poglitsch}, {Geis}, {Stacey}, \&
  {Townes}}]{Madden1997}
{Madden}, S.~C., {Poglitsch}, A., {Geis}, N., {Stacey}, G.~J., \& {Townes},
  C.~H. 1997, \apj, 483, 200

\bibitem[{{Madden} {et~al.}(2020){Madden}, {Cormier}, {Hony}, {Lebouteiller},
  {Abel}, {Galametz}, {De Looze}, {Chevance}, {Polles}, {Lee}, {Galliano},
  {Lambert-Huyghe}, {Hu}, \& {Ramambason}}]{Madden2020}
{Madden}, S.~C., {Cormier}, D., {Hony}, S., {et~al.} 2020, \aap, 643, A141

\bibitem[{{Makiuti} {et~al.}(2002){Makiuti}, {Shibai}, {Nakagawa}, {Okuda},
  {Okumura}, {Matsuhara}, {Hiromoto}, \& {Doi}}]{Makiuti2002}
{Makiuti}, S., {Shibai}, H., {Nakagawa}, T., {et~al.} 2002, \aap, 382, 600

\bibitem[{Mendenhall \& Sincich(2011)}]{mendenhall}
Mendenhall, W., \& Sincich, T. 2011, A Second Course in Statistics: Regression
  Analysis (Pearson Education)

\bibitem[{{Meyer} {et~al.}(1997){Meyer}, {Cardelli}, \& {Sofia}}]{Meyer1997}
{Meyer}, D.~M., {Cardelli}, J.~A., \& {Sofia}, U.~J. 1997, \apjl, 490, L103

\bibitem[{{Mookerjea} {et~al.}(2016){Mookerjea}, {Israel}, {Kramer}, {Nikola},
  {Braine}, {Ossenkopf}, {R{\"o}llig}, {Henkel}, {van der Werf}, {van der Tak},
  \& {Wiedner}}]{Mookerjea2016}
{Mookerjea}, B., {Israel}, F., {Kramer}, C., {et~al.} 2016, \aap, 586, A37

\bibitem[{{Mu{\~n}oz-Mateos} {et~al.}(2009){Mu{\~n}oz-Mateos}, {Gil de Paz},
  {Zamorano}, {Boissier}, {Dale}, {P{\'e}rez-Gonz{\'a}lez}, {Gallego},
  {Madore}, {Bendo}, {Boselli}, {Buat}, {Calzetti}, {Moustakas}, \&
  {Kennicutt}}]{Munoz2009}
{Mu{\~n}oz-Mateos}, J.~C., {Gil de Paz}, A., {Zamorano}, J., {et~al.} 2009,
  \apj, 703, 1569

\bibitem[{{Murphy} {et~al.}(2018){Murphy}, {Khan}, {Williams}, {Dolphin},
  {Dalcanton}, \& {D{\'\i}az-Rodr{\'\i}guez}}]{Murphy2018}
{Murphy}, J.~W., {Khan}, R., {Williams}, B., {et~al.} 2018, \apj, 860, 117

\bibitem[{{Nicholls} {et~al.}(2017){Nicholls}, {Sutherland}, {Dopita},
  {Kewley}, \& {Groves}}]{Nicholls2017}
{Nicholls}, D.~C., {Sutherland}, R.~S., {Dopita}, M.~A., {Kewley}, L.~J., \&
  {Groves}, B.~A. 2017, \mnras, 466, 4403

\bibitem[{{Nieva} \& {Przybilla}(2012)}]{Nieva2012}
{Nieva}, M.~F., \& {Przybilla}, N. 2012, \aap, 539, A143

\bibitem[{{Oberst} {et~al.}(2006){Oberst}, {Parshley}, {Stacey}, {Nikola},
  {L{\"o}hr}, {Harnett}, {Tothill}, {Lane}, {Stark}, \& {Tucker}}]{Oberst2006}
{Oberst}, T.~E., {Parshley}, S.~C., {Stacey}, G.~J., {et~al.} 2006, \apjl, 652,
  L125

\bibitem[{{Okada} {et~al.}(2019){Okada}, {G{\"u}sten}, {Requena-Torres},
  {R{\"o}llig}, {Stutzki}, {Graf}, \& {Hughes}}]{okada2019}
{Okada}, Y., {G{\"u}sten}, R., {Requena-Torres}, M.~A., {et~al.} 2019, \aap,
  621, A62

\bibitem[{{Okada} {et~al.}(2015){Okada}, {Requena-Torres}, {G{\"u}sten},
  {Stutzki}, {Wiesemeyer}, {P{\"u}tz}, \& {Ricken}}]{okada2015}
{Okada}, Y., {Requena-Torres}, M.~A., {G{\"u}sten}, R., {et~al.} 2015, \aap,
  580, A54

\bibitem[{{Ostriker} {et~al.}(2010){Ostriker}, {McKee}, \&
  {Leroy}}]{ostriker2010}
{Ostriker}, E.~C., {McKee}, C.~F., \& {Leroy}, A.~K. 2010, \apj, 721, 975

\bibitem[{{P{\'e}rez-Beaupuits} {et~al.}(2015){P{\'e}rez-Beaupuits}, {Stutzki},
  {Ossenkopf}, {Spaans}, {G{\"u}sten}, \& {Wiesemeyer}}]{perez2015}
{P{\'e}rez-Beaupuits}, J.~P., {Stutzki}, J., {Ossenkopf}, V., {et~al.} 2015,
  \aap, 575, A9

\bibitem[{{Pety}(2005)}]{2005sf2a.conf..721P}
{Pety}, J. 2005, in SF2A-2005: Semaine de l'Astrophysique Francaise, ed.
  F.~{Casoli}, T.~{Contini}, J.~M. {Hameury}, \& L.~{Pagani}, 721

\bibitem[{{Pilyugin} {et~al.}(2014){Pilyugin}, {Grebel}, {Zinchenko}, \&
  {Kniazev}}]{Pilyugin2014}
{Pilyugin}, L.~S., {Grebel}, E.~K., {Zinchenko}, I.~A., \& {Kniazev}, A.~Y.
  2014, \aj, 148, 134

\bibitem[{{Pineda} {et~al.}(2014){Pineda}, {Langer}, \&
  {Goldsmith}}]{Pineda2014}
{Pineda}, J.~L., {Langer}, W.~D., \& {Goldsmith}, P.~F. 2014, \aap, 570, A121

\bibitem[{{Pineda} {et~al.}(2013){Pineda}, {Langer}, {Velusamy}, \&
  {Goldsmith}}]{pineda2013}
{Pineda}, J.~L., {Langer}, W.~D., {Velusamy}, T., \& {Goldsmith}, P.~F. 2013,
  \aap, 554, A103

\bibitem[{{Pineda} {et~al.}(2017){Pineda}, {Langer}, {Goldsmith}, {Horiuchi},
  {Kuiper}, {Muller}, {Hughes}, {Ott}, {Requena-Torres}, {Velusamy}, \&
  {Wong}}]{Pineda2017}
{Pineda}, J.~L., {Langer}, W.~D., {Goldsmith}, P.~F., {et~al.} 2017, \apj, 839,
  107

\bibitem[{{Requena-Torres} {et~al.}(2016){Requena-Torres}, {Israel}, {Okada},
  {G{\"u}sten}, {Stutzki}, {Risacher}, {Simon}, \& {Zinnecker}}]{requena2016}
{Requena-Torres}, M.~A., {Israel}, F.~P., {Okada}, Y., {et~al.} 2016, \aap,
  589, A28

\bibitem[{{Risacher} {et~al.}(2016){Risacher}, {G{\"u}sten}, {Stutzki},
  {H{\"u}bers}, {Bell}, {Buchbender}, {B{\"u}chel}, {Csengeri}, {Graf},
  {Heyminck}, {Higgins}, {Honingh}, {Jacobs}, {Klein}, {Okada}, {Parikka},
  {P{\"u}tz}, {Reyes}, {Ricken}, {Riquelme}, {Simon}, \&
  {Wiesemeyer}}]{Risacher2016}
{Risacher}, C., {G{\"u}sten}, R., {Stutzki}, J., {et~al.} 2016, \aap, 595, A34

\bibitem[{{Risacher} {et~al.}(2018){Risacher}, {G{\"u}sten}, {Stutzki},
  {H{\"u}bers}, {Aladro}, {Bell}, {Buchbender}, {B{\"u}chel}, {Csengeri},
  {Duran}, {Graf}, {Higgins}, {Honingh}, {Jacobs}, {Justen}, {Klein},
  {Mertens}, {Okada}, {Parikka}, {P{\"u}tz}, {Reyes}, {Richter}, {Ricken},
  {Riquelme}, {Rothbart}, {Schneider}, {Simon}, {Wienold}, {Wiesemeyer},
  {Ziebart}, {Fusco}, {Rosner}, \& {Wohler}}]{Risacher2018}
---. 2018, Journal of Astronomical Instrumentation, 7, 1840014

\bibitem[{{R{\"o}llig} {et~al.}(2016){R{\"o}llig}, {Simon}, {G{\"u}sten},
  {Stutzki}, {Israel}, \& {Jacobs}}]{rollig2016}
{R{\"o}llig}, M., {Simon}, R., {G{\"u}sten}, R., {et~al.} 2016, \aap, 591, A33

\bibitem[{{Schlegel} {et~al.}(1998){Schlegel}, {Finkbeiner}, \&
  {Davis}}]{Schlegel1998}
{Schlegel}, D.~J., {Finkbeiner}, D.~P., \& {Davis}, M. 1998, \apj, 500, 525

\bibitem[{{Sch{\"o}ier} {et~al.}(2005){Sch{\"o}ier}, {van der Tak}, {van
  Dishoeck}, \& {Black}}]{schoier2005}
{Sch{\"o}ier}, F.~L., {van der Tak}, F.~F.~S., {van Dishoeck}, E.~F., \&
  {Black}, J.~H. 2005, \aap, 432, 369

\bibitem[{{Shibai} {et~al.}(1991){Shibai}, {Okuda}, {Nakagawa}, {Matsuhara},
  {Maihara}, {Mizutani}, {Kobayashi}, {Hiromoto}, {Nishimura}, \&
  {Low}}]{Shibai1991}
{Shibai}, H., {Okuda}, H., {Nakagawa}, T., {et~al.} 1991, \apj, 374, 522

\bibitem[{{Sim{\'o}n-D{\'\i}az} \& {Stasi{\'n}ska}(2011)}]{Simon-Diaz2011}
{Sim{\'o}n-D{\'\i}az}, S., \& {Stasi{\'n}ska}, G. 2011, \aap, 526, A48

\bibitem[{{Smith} {et~al.}(2017){Smith}, {Croxall}, {Draine}, {De Looze},
  {Sandstrom}, {Armus}, {Beir{\~a}o}, {Bolatto}, {Boquien}, {Brandl},
  {Crocker}, {Dale}, {Galametz}, {Groves}, {Helou}, {Herrera-Camus}, {Hunt},
  {Kennicutt}, {Walter}, \& {Wolfire}}]{Smith2017}
{Smith}, J.~D.~T., {Croxall}, K., {Draine}, B., {et~al.} 2017, \apj, 834, 5

\bibitem[{{Sofia} {et~al.}(2004){Sofia}, {Lauroesch}, {Meyer}, \&
  {Cartledge}}]{Sofia2004}
{Sofia}, U.~J., {Lauroesch}, J.~T., {Meyer}, D.~M., \& {Cartledge}, S. I.~B.
  2004, \apj, 605, 272

\bibitem[{{Stacey} {et~al.}(1991){Stacey}, {Geis}, {Genzel}, {Lugten},
  {Poglitsch}, {Sternberg}, \& {Townes}}]{Stacey1991}
{Stacey}, G.~J., {Geis}, N., {Genzel}, R., {et~al.} 1991, \apj, 373, 423

\bibitem[{{Stacey} {et~al.}(1985){Stacey}, {Viscuso}, {Fuller}, \&
  {Kurtz}}]{Stacey1985}
{Stacey}, G.~J., {Viscuso}, P.~J., {Fuller}, C.~E., \& {Kurtz}, N.~T. 1985,
  \apj, 289, 803

\bibitem[{{Sutter} {et~al.}(2019){Sutter}, {Dale}, {Croxall}, {Pelligrini},
  {Smith}, {Appleton}, {Beir{\~a}o}, {Bolatto}, {Calzetti}, {Crocker}, {De
  Looze}, {Draine}, {Galametz}, {Groves}, {Helou}, {Herrera-Camus}, {Hunt},
  {Kennicutt}, {Roussel}, \& {Wolfire}}]{Sutter2019}
{Sutter}, J., {Dale}, D.~A., {Croxall}, K.~V., {et~al.} 2019, \apj, 886, 60

\bibitem[{{Tayal}(2008)}]{Tayal2008}
{Tayal}, S.~S. 2008, \aap, 486, 629

\bibitem[{{Tayal}(2011)}]{Tayal2011}
---. 2011, \apjs, 195, 12

\bibitem[{{The Astropy Collaboration} {et~al.}(2018){The Astropy
  Collaboration}, {Price-Whelan}, {Sip{\H o}cz}, {G{\"u}nther}, {Lim},
  {Crawford}, {Conseil}, {Shupe}, {Craig}, {Dencheva}, {Ginsburg},
  {VanderPlas}, {Bradley}, {P{\'e}rez-Su{\'a}rez}, {de Val-Borro}, {Aldcroft},
  {Cruz}, {Robitaille}, {Tollerud}, {Ardelean}, {Babej}, {Bachetti}, {Bakanov},
  {Bamford}, {Barentsen}, {Barmby}, {Baumbach}, {Berry}, {Biscani}, {Boquien},
  {Bostroem}, {Bouma}, {Brammer}, {Bray}, {Breytenbach}, {Buddelmeijer},
  {Burke}, {Calderone}, {Cano Rodr{\'{\i}}guez}, {Cara}, {Cardoso},
  {Cheedella}, {Copin}, {Crichton}, {D{\'A}vella}, {Deil}, {Depagne},
  {Dietrich}, {Donath}, {Droettboom}, {Earl}, {Erben}, {Fabbro}, {Ferreira},
  {Finethy}, {Fox}, {Garrison}, {Gibbons}, {Goldstein}, {Gommers}, {Greco},
  {Greenfield}, {Groener}, {Grollier}, {Hagen}, {Hirst}, {Homeier}, {Horton},
  {Hosseinzadeh}, {Hu}, {Hunkeler}, {Ivezi{\'c}}, {Jain}, {Jenness}, {Kanarek},
  {Kendrew}, {Kern}, {Kerzendorf}, {Khvalko}, {King}, {Kirkby}, {Kulkarni},
  {Kumar}, {Lee}, {Lenz}, {Littlefair}, {Ma}, {Macleod}, {Mastropietro},
  {McCully}, {Montagnac}, {Morris}, {Mueller}, {Mumford}, {Muna}, {Murphy},
  {Nelson}, {Nguyen}, {Ninan}, {N{\"o}the}, {Ogaz}, {Oh}, {Parejko}, {Parley},
  {Pascual}, {Patil}, {Patil}, {Plunkett}, {Prochaska}, {Rastogi}, {Reddy
  Janga}, {Sabater}, {Sakurikar}, {Seifert}, {Sherbert}, {Sherwood-Taylor},
  {Shih}, {Sick}, {Silbiger}, {Singanamalla}, {Singer}, {Sladen}, {Sooley},
  {Sornarajah}, {Streicher}, {Teuben}, {Thomas}, {Tremblay}, {Turner},
  {Terr{\'o}n}, {van Kerkwijk}, {de la Vega}, {Watkins}, {Weaver}, {Whitmore},
  {Woillez}, \& {Zabalza}}]{astropy}
{The Astropy Collaboration}, {Price-Whelan}, A.~M., {Sip{\H o}cz}, B.~M.,
  {et~al.} 2018, ArXiv e-prints, arXiv:1801.02634

\bibitem[{{Tielens} \& {Hollenbach}(1985)}]{Tielens1985}
{Tielens}, A.~G.~G.~M., \& {Hollenbach}, D. 1985, \apj, 291, 722

\bibitem[{{Velusamy} {et~al.}(2017){Velusamy}, {Langer}, {Goldsmith}, \&
  {Pineda}}]{Velusamy2017}
{Velusamy}, T., {Langer}, W.~D., {Goldsmith}, P.~F., \& {Pineda}, J.~L. 2017,
  \apj, 838, 165

\bibitem[{{Walter} {et~al.}(2008){Walter}, {Brinks}, {de Blok}, {Bigiel},
  {Kennicutt}, {Thornley}, \& {Leroy}}]{Walter2008}
{Walter}, F., {Brinks}, E., {de Blok}, W.~J.~G., {et~al.} 2008, \aj, 136, 2563

\bibitem[{{Welty} {et~al.}(2016){Welty}, {Lauroesch}, {Wong}, \&
  {York}}]{Welty2016}
{Welty}, D.~E., {Lauroesch}, J.~T., {Wong}, T., \& {York}, D.~G. 2016, \apj,
  821, 118

\bibitem[{{Wolfire} {et~al.}(2010){Wolfire}, {Hollenbach}, \&
  {McKee}}]{Wolfire2010}
{Wolfire}, M.~G., {Hollenbach}, D., \& {McKee}, C.~F. 2010, \apj, 716, 1191

\bibitem[{{Wolfire} {et~al.}(1995){Wolfire}, {Hollenbach}, {McKee}, {Tielens},
  \& {Bakes}}]{wolfire1995}
{Wolfire}, M.~G., {Hollenbach}, D., {McKee}, C.~F., {Tielens}, A.~G.~G.~M., \&
  {Bakes}, E.~L.~O. 1995, \apj, 443, 152

\bibitem[{{Wolfire} {et~al.}(1989){Wolfire}, {Hollenbach}, \&
  {Tielens}}]{Wolfire1989}
{Wolfire}, M.~G., {Hollenbach}, D., \& {Tielens}, A.~G.~G.~M. 1989, \apj, 344,
  770

\bibitem[{{Wolfire} {et~al.}(2003){Wolfire}, {McKee}, {Hollenbach}, \&
  {Tielens}}]{Wolfire2003}
{Wolfire}, M.~G., {McKee}, C.~F., {Hollenbach}, D., \& {Tielens}, A.~G.~G.~M.
  2003, \apj, 587, 278

\bibitem[{{Wright} {et~al.}(1991){Wright}, {Mather}, {Bennett}, {Cheng},
  {Shafer}, {Fixsen}, {Eplee}, {Isaacman}, {Read}, {Boggess}, {Gulkis},
  {Hauser}, {Janssen}, {Kelsall}, {Lubin}, {Meyer}, {Moseley}, {Murdock},
  {Silverberg}, {Smoot}, {Weiss}, \& {Wilkinson}}]{Wright1991}
{Wright}, E.~L., {Mather}, J.~C., {Bennett}, C.~L., {et~al.} 1991, \apj, 381,
  200

\bibitem[{{Zanella} {et~al.}(2018){Zanella}, {Daddi}, {Magdis}, {Diaz Santos},
  {Cormier}, {Liu}, {Cibinel}, {Gobat}, {Dickinson}, {Sargent}, {Popping},
  {Madden}, {Bethermin}, {Hughes}, {Valentino}, {Rujopakarn}, {Pannella},
  {Bournaud}, {Walter}, {Wang}, {Elbaz}, \& {Coogan}}]{Zanella2018}
{Zanella}, A., {Daddi}, E., {Magdis}, G., {et~al.} 2018, \mnras, 481, 1976

\end{thebibliography}

\end{document}